\documentclass[amsmath,amssymb,aps]{revtex4}

\selectfont 

\usepackage{amsmath}
\usepackage{graphicx}
\usepackage{bm}
\usepackage{braket}
\usepackage{tensor}
\usepackage{color}
\usepackage{slashed}
\usepackage{color}
\usepackage{booktabs} % to make nice table
\usepackage{multirow}
\usepackage{float}

\def\omegabar{\overline{\omega}}

\newcommand{\beq}{\begin{equation}}
\newcommand{\eeq}{\end{equation}}
\newcommand{\beqy}{\begin{eqnarray}}
\newcommand{\eeqy}{\end{eqnarray}}
\def\be{\begin{equation}}
\def\ee{\end{equation}}
\def\bea{\begin{eqnarray}}
\def\eea{\end{eqnarray}}

\newcommand{\Lag}{\mathcal{L}}

\begin{document}

\title{Charged scalar perturbations on charged black holes in dRGT massive gravity}

\author{Piyabut Burikham}
\email{piyabut@gmail.com}

\affiliation{High Energy Physics Theory Group, Department of Physics, Faculty of Science,
Chulalongkorn University, Phyathai Rd., Bangkok 10330, Thailand}

\author{Supakchai Ponglertsakul}
\email{supakchai.p@gmail.com}

\affiliation{Theoretical and Computational Physics Group, Theoretical and Computational Science Center(TaCS), Faculty of Science,
King Mongkut's University of Technology Thonburi, Prachautid Rd., Bangkok 10140, Thailand}

\author{Lunchakorn Tannukij}
\email{l_tannukij@hotmail.com}

\affiliation{Department of Physics, Faculty of Science,
Mahidol University, Bangkok 10400, Thailand}
\affiliation{Department of Physics, Hanyang University, Seoul 133-891, South Korea}

\date{\today}

\begin{abstract}
We explore the quasi-stationary profile of massive charged scalar field in a class of charged black hole in dRGT massive gravity.  We discuss how the linear term in the metric which is a unique character of the dRGT massive gravity affects structure of the spacetime. Numerical calculations of the quasinormal modes are performed for the charged scalar field in the dRGT black hole background. For asymptotically de Sitter~(dS) black hole, an improved asymptotic iteration method is used to obtain the associated quasinormal frequencies. The unstable modes are found for $\ell=0$ case and their corresponding real parts satisfy superradiant condition. For $\ell=2$, the results show that all the de Sitter black holes considered here are stable against a small perturbation. For asymptotically dRGT anti de Sitter~(AdS) black hole, unstable modes are found with the frequency satisfying superradiant condition. Effects of massive gravity parameter are discussed. Analytic calculation reveals unique diffusive nature of quasinormal modes in the massive gravity model with the linear term.  Numerical results confirm existence of the characteristic diffusive modes in both dS and AdS cases.

\end{abstract}

\pacs{nnnnn}

\maketitle

%\section{Introduction}\label{sec:intro}
%
%Superradiant instability...
%
%In this paper....
%
%The outline of this paper is as follows.

\section{Introduction}\label{sec:intro}

Massive gravity is a modified gravity theory in which gravity is described by a massive spin-2 graviton, propagating 5 degrees of freedom. Unlike general relativity whose graviton is massless, the gravitation in massive gravity is essentially modified at the scale corresponding to the graviton mass $m_g$. In cosmological aspects, we might expect this characteristic to be responsible for the cosmic accelerating expansion given that the graviton mass is of the same order as the Hubble parameter; $m_g \sim H \sim 10^{-33}$ eV \cite{Bonvin:2016crt}. On the other hand, the recent observation from the Laser Interferometer Gravitational-Wave Observatory (LIGO) on binary black hole merger, GW150914, has put an upper bound $m_g \leq 1.2 \times 10^{-22}$ eV for the graviton mass \cite{TheLIGOScientific:2016src} (see also \cite{deRham:2016nuf} for graviton mass bounds from other aspects). In cosmological point of view, massive gravity is still a viable model of the universe.

The very first model of massive gravity was realized as a linear theory by Fierz and Pauli (FP) in 1939 \cite{Fierz:1939ix}. The FP massive gravity was then proven that the theory suffers from the van Dam-Veltman-Zakharov (vDVZ) discontinuity where the predictions made by the FP theory do not coincide with those made by general relativity when an appropriate limit (massless-graviton limit) is taken \cite{vanDam:1970vg,Zakharov:1970cc}. After that, Vainshtein suggested that because of the introduction of the graviton mass, the graviton mass introduces a new scale known as Vainshtein radius outside which the FP theory works with good accuracy \cite{Vainshtein:1972sx}. For the massless limit, however, this scale is pushed towards infinity so that the linear theory cannot be trusted when being used for local systems and nonlinear effects should be included in order to cure the vDVZ discontinuity \cite{Vainshtein:1972sx}. However, it was found by Boulware and Deser that generic nonlinear massive gravity theories always propagate 6 degrees of freedom instead of 5 and the additional degree of freedom unfortunately has wrong-sign kinetic energy (known as BD ghost), causing an instability to the theories \cite{Boulware:1973my}. In 2010, de Rham, Gabadadze, and Tolley found that there exists a class of nonlinear massive gravity theory which does not possess the BD ghost, dubbed dRGT massive gravity \cite{deRham:2010ik,deRham:2010kj}. Since this theory is constructed successfully without the well-known pathology, it actually gives rise to various kinds of studies in massive gravity such as cosmological solutions \cite{D'Amico:2011jj,Gumrukcuoglu:2011ew}, cosmological perturbations \cite{Gumrukcuoglu:2011zh}, black hole solutions and thermodynamics \cite{Cai:2014znn,Ghosh:2015cva}, and even various generalizations of the dRGT theory, like the quasi-dilaton theory \cite{DeFelice:2013za,DeFelice:2013gm}. Recently, a black hole solution to dRGT massive gravity has been found \cite{Cai:2014znn,Ghosh:2015cva} and the solution is in agreement with the dRGT cosmology in that the graviton mass effectively plays a role of cosmological constant. It was also found that the solution can be stable in the thermodynamics language \cite{Cai:2014znn,Ghosh:2015cva}. 

A bosonic field can be used to extract rotational energy and electric charge from a black hole via the so-called superradiant scattering. If the frequency of the bosonic field on the black hole spacetime satisfies the following (for asymptotically flat spacetime) \cite{Bekenstein:1973mi}

\beq
\omega < m\Omega_H + q \Phi_H,
\eeq 

where $m$ is azimuthal number, $q$ is particle charge, $\Omega_H$ and $\Phi_H$ are angular velocity and electrostatic potential at the black hole horizon respectively. The superradiant phenomena can often lead to an instability of the spacetime background if superradiant mode is confined near the black hole horizon. The amplitude of the bosonic field will be amplified repeatedly causing a non-negligible back-reaction on the exterior geometry. 

In standard general relativity, complex scalar field on Reissner-Nordstr\"om (RN) background is known to be suffered from superradiant instability. For example, massive charged scalar field on RN enclosed with a mirror-like boundary condition experiences charged superradiant instability \cite{Herdeiro:2013pia}. Time domain analysis \cite{Degollado:2013bha} on this system reveals that the unstable modes grow a lot faster than in the rotating case. Moreover, a massless charged scalar field on a small RN black hole in asymptotically anti-de Sitter (AdS) is shown to be superradiantly unstable  \cite{Uchikata:2011zz}. Despite RN in asymptotically flat spacetime is stable against spherically symmetric charged scalar perturbations, however, an instability of the RN black hole in asymptotically de Sitter spacetime is surprisingly discovered \cite{Zhu:2014sya}. It is shown in \cite{Konoplya:2014lha} that instability occurs when the scalar field's frequency satisfies the superradiant condition. It should be noted that not all the superradiant modes are unstable, the instability holds only for spherical perturbation $\ell=0$ mode while the superradiant mode exists in higher $\ell$.

A new class of an exact spherically symmetric neutral/charged black hole solutions in dRGT massive gravity are found in \cite{Ghosh:2015cva}. The effective cosmological constant naturally arises in the theory and can be written in term of the graviton mass. One could treat these black holes either as modified Schwarzschild/Reissner-Nordstr\"om with positive or negative cosmological constant depending on the choice of free parameters. In addition, scalar perturbation on neutral/charged dRGT black holes and their thermodynamic behaviour are studied in \cite{Prasia:2016fcc}. A natural question that one might ask is whether these dRGT black holes experience superradiant phenomena. Could dS and AdS boundary lead to an instability caused by the superradiant effect? What is the effect of massive charged scalar field on the charged dRGT black holes in asymptotically dS and AdS spacetimes?

The main purpose of this paper is to study the perturbation of massive charged scalar field in the dRGT black hole spacetime. This is equivalent to the study of quasinormal modes (QNMs) of black holes in the scalar channel, with extension to the complex scalar perturbations. In contrast to the normal modes, QNMs decay/grow with complex frequencies which are uniquely determined by black hole's physical parameters i.e., mass, charge and angular momentum. Existence of the unique linear term in the metric of the dRGT model inevitably alters the QNMs of the charged scalar in such background. We address such behaviour in this paper. In section \ref{sec:formal}, we introduce the basic set-up for constructing charged black hole solution in dRGT massive gravity. Most of the details discussed in this section originates from the work done in \cite{Ghosh:2015cva}. Then we discuss the effects of linear term $(\gamma)$ which is the unique character of the black holes in dRGT massive gravity in section \ref{sec:gamma}. In section \ref{sec:waveEq}, the Klein-Gordon equation of massive charged scalar field on the dRGT black hole spacetime is derived. Then the QNMs of dRGT black holes with a positive cosmological constant are explored in section \ref{sec:dS}. The QNMs of dRGT black holes with a negative cosmological constant are calculated in section \ref{sec:AdS}. In section \ref{sec:Ana}, we provide an analytic calculation for the diffusive modes (QNMs with zero real part) of the dRGT background. Our conclusions are presented in section \ref{sec:conclude}.

\section{Formalism}\label{sec:formal}

The dRGT massive gravity coupled with massive charged scalar field is described by the following action \cite{deRham:2010kj} (with $c=8\pi G=1$). 
\begin{align}
S &= \frac{1}{2}\int d^4 x \sqrt{-g}\left[R + m^2_g \mathcal{U}(g,\phi^a) + \mathcal{L}_m \right], \label{action}
\end{align}
where the matter Lagrangian is 
\begin{align}
\mathcal{L}_m &\equiv \mathcal{L}_{EM} + \mathcal{L}_{\Phi}, \nonumber \\
&= - \frac{1}{2}F_{\mu\nu}F^{\mu\nu} - g^{\mu\nu}D^\ast_{(\mu} \Phi^\ast D^{}_{\nu)} \Phi - m_s^2\Phi^\ast\Phi.
\end{align}
Graviton mass and scalar field mass are denoted by $m_g$ and $m_s$ respectively. The symmetrized combination of indices is defined as $X_{(\mu\nu)}=\frac{1}{2}\left(X_{\mu\nu}+X_{\nu\mu}\right)$. The field strength tensor in the curved spacetime is given by $F_{\mu\nu} = A_{\nu;\mu} - A_{\mu;\nu}$ and the covariant derivative in the presence of the gauge symmetry is $D_\mu = \nabla_\mu - i q A_\mu$ where $A_\mu$ is the electromagnetic potential and $q$ is charge of the scalar field $\Phi$. 

The ghost-free massive graviton self-interacting potential is given by 
\begin{align}
\mathcal{U}(g,\phi^a) &= \mathcal{U}_2 + \alpha_3 \mathcal{U}_3 + \alpha_4 \mathcal{U}_4
\end{align}
where
\begin{align}
\alpha_3 &= \frac{\alpha-1}{3}, \\
\alpha_4 &= \frac{\beta}{4} + \frac{1-\alpha}{12}. \\
\mathcal{U}_2 &= [\mathcal{K}]^2 - [\mathcal{K}^2], \\
\mathcal{U}_3 &= [\mathcal{K}]^3 - 3[\mathcal{K}][\mathcal{K}^2] + 2[\mathcal{K}^3], \\
\mathcal{U}_4 &= [\mathcal{K}]^4 - 6[\mathcal{K}]^2[\mathcal{K}^2] + 8[\mathcal{K}][\mathcal{K}^3] + 3[\mathcal{K}^2]^2 - 6[\mathcal{K}^4].
\end{align}
$\alpha$ and $\beta$ are free parameters. $K^{\mu}_{\nu} = \delta^{\mu}_{\nu} - \sqrt{g^{\mu\sigma}f_{ab}\partial_{\sigma}\phi^a\partial_{\nu}\phi^b}$. $[\mathcal{K}]=\mathcal{K}^{\mu}_{\mu}$ and $[\mathcal{K}^n]=(\mathcal{K}^n)^{\mu}_{\mu}$. We will work in the unitary gauge for which the four St\"{u}ckelberg fields take the form $\phi^a = x^{\mu}\delta^a_{\mu}$. The fiducial metric is chosen to be $f_{ab} = diag(0,0,c^2,c^2\sin^2\theta)$, where $c$ is a constant. 

\subsection{Field equations}

By varying (\ref{action}), three equations of motions are obtained
\begin{align}
R_{\mu\nu} - \frac{1}{2} R g_{\mu\nu} &= -m^2_g X_{\mu\nu} +  \left( T_{\mu\nu}^{F} + T_{\mu\nu}^\Phi \right) \label{einstein-eq}, \\
\tensor{F}{^{\mu\nu} _{; \mu}} &= J^\nu, \\
D_a D^a \Phi &=  m_s^2 \Phi, \label{scalar-eq}
\end{align}
where $X_{\mu\nu}$ is given by \cite{Ghosh:2015cva}

\begin{align}
 X_{\mu\nu} &= {\cal K}_ {\mu\nu} -{\cal K}g_ {\mu\nu} -\alpha\left\{{\cal K}^2_{\mu\nu}-{\cal K}{\cal K}_{\mu\nu} +\frac{[{\cal K}]^2-[{\cal K}^2]}{2}g_{\mu\nu}\right\} \nonumber\\
  & +3\beta\left\{ {\cal K}^3_{\mu\nu} -{\cal K}{\cal K}^2_{\mu\nu} +\frac{1}{2}{\cal K}_{\mu\nu}\left\{[{\cal K}]^2 -[{\cal K}^2]\right\} \right.\nonumber
  \\
 & \left. - \frac{1}{6}g_{\mu\nu}\left\{[{\cal K}]^3 -3[{\cal K}][{\cal K}^2] +2[{\cal K}^3]\right\} \right\} . \label{effemt}
\end{align}
The energy-momentum tensor of the gauge and scalar field are 
\begin{align}
T_{\mu\nu}^{F} &= F_{\mu \gamma} \tensor{F}{_\nu ^\gamma} - \frac{1}{4} g_{\mu\nu} F_{\gamma\lambda} F^{\gamma\lambda} \\
T_{\mu\nu}^{\Phi} &= D^\ast_{(\mu} \Phi^\ast D^{}_{\nu)} \Phi + g_{\mu\nu} \Lag_\Phi.
\end{align}
Finally, the Noether current $J^\nu$ of the scalar field is 
\begin{align}
J^{\nu} &=  \frac{iq}{2} \left(\Phi^\ast D^\nu \Phi - \Phi (D^\nu \Phi)^\ast \right).
\end{align}

\subsection{Black hole solutions}
In the absence of charged scalar field $T^{\Phi}_{\mu\nu}=0$, the Einstein's equations (\ref{einstein-eq}) admits a static spherically symmetric solution in the following form \cite{Ghosh:2015cva}
\begin{align}
ds^{2} &= -f(r)dt^2 + f^{-1}dr^2 + r^2d\theta^2 + r^2\sin^2\theta d\varphi^2, \label{BHmetric}
\end{align}
where 
\begin{align}
f(r) &= 1 - \frac{2M}{r} + \frac{Q^2}{r^2} - \frac{\Lambda}{3}r^2 + \gamma r + \epsilon, \label{metric} \\
\Lambda &= -3m^2_g(1+\alpha+\beta), \label{lamb} \\
\gamma &= -cm^2_g(1+2\alpha+3\beta), \label{gam} \\
\epsilon &= c^2m^2_g(\alpha+3\beta).
\end{align}
The mass and electric charge of the black hole are denoted by $M$ and $Q$ respectively where $\epsilon$ is a constant. If $(1+\alpha+\beta) > 0$, we obtain modified Reissner-Nordstr\"om-AdS solution while $(1+\alpha+\beta) < 0$ yields the modified dS-type solution. In the limit $c\rightarrow0$ which sets $\gamma=\epsilon=0$, the metric (\ref{metric}) becomes the standard Reissner-Nordstr\"om solution with a cosmological constant. In addition, if the graviton mass is set to zero, we obtain the asymptotically flat Reissner-Nordstr\"om solution.

Apart from the parameters $\alpha$ and $\beta$ paramatrising cubic and quartic graviton interactions, there are two main effects of massive gravity in dRGT model reflected in two parameters: the graviton mass $m_{g}$ and the parameter $c$ in the fiducial metric $f_{ab}=diag(0,0,c^{2},c^{2}\sin^{2}\theta)$.  After setting $\epsilon$ to zero \cite{1stfootnote}, all physical parameters in the metric (\ref{metric}) depend on $m_{g}^{2}$ but only $\gamma$, presenting linear term in $r$, depends on $c$.  The cosmological constant $\Lambda$, on the other hand, does not depend on the fiducial metric parameter $c$.  The two quantities $\Lambda$ and $\gamma$ are thus independent.  We can have massive gravity model with $m_{g}\neq 0$ but vanishing $c$ which will lead to only cosmological constant term in the metric.  Or we can have massive gravity model with nonzero $c$ resulting in the existence of linear term $\gamma r$ in the metric in addition to the cosmological constant term.

\section{Effects of $\gamma$ parameter}\label{sec:gamma}

We will consider effects of the $\gamma$ term unique in massive gravity model in this section.  It will be shown for fixed physical parameters $M, Q, \Lambda$ and $\epsilon=0$, that varying $\gamma$ could lead to spacetime with differing properties starting from regular spacetime to black hole and extremal black hole.  This is unique to the spacetime in massive gravity theories.  Since we expect flat spacetime with usual radial coordinate centered at $r=0$ due to spherical symmetry, it is reasonable to set $\epsilon=0$ implying $\alpha= -3\beta$.  We are thus left with 2 independent parameters $\beta$ and $c$.  For a fixed value of graviton mass $m_{g}^{2}$, $\Lambda$ and $\gamma$ given by (\ref{lamb}) and (\ref{gam}) remain independent.

%Since when $r=0$, we expect flat spacetime with usual radial coordinate due to spherical symmetry  

\subsection{Positive $\Lambda$}

\begin{figure}
\centering
 \includegraphics[width=0.7\textwidth]{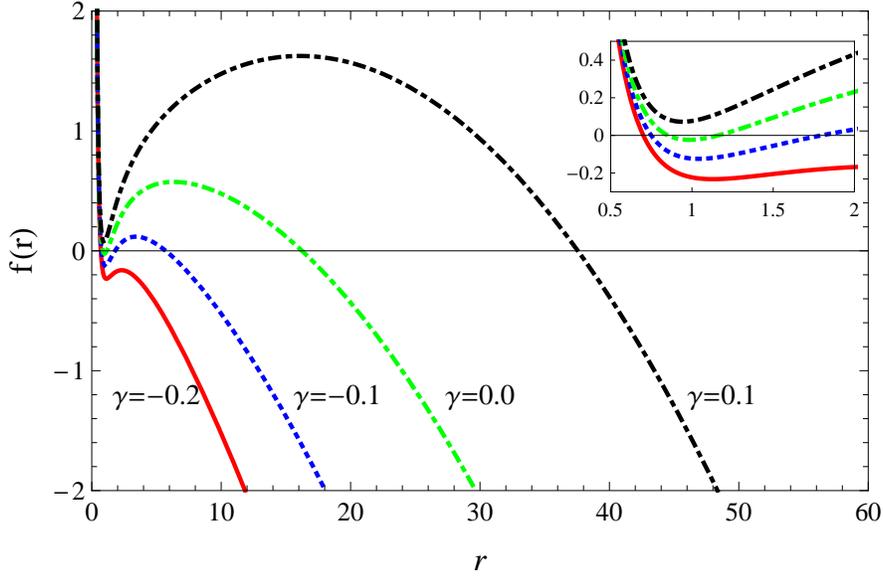}
\caption{The behaviour of metric function $f(r)$ plots against radius for various values of $\gamma$ with fixed $M=1,Q=0.99,\Lambda=0.01,\epsilon=0$. A subplot shows the behaviour of $f(r)$ when $r$ is small. }
 \label{fig:dsfig1}
\end{figure}

In general, the metric function (\ref{metric}) has four roots. It is possible that all the roots are real. More specifically, for dS-type solution, there will be three positive roots and one negative root. All three positive roots will be treated as Cauchy horizon $r_m$, event horizon $r_h$ and cosmological horizon $r_c$ where $r_m<r_h<r_c$. The root structure of metric function (\ref{metric}) is shown in Fig.~\ref{fig:dsfig1}. In this plot, we fix the black hole mass $M$, charge $Q$, cosmological constant $\Lambda$ and $\epsilon$ to be $M=1, Q=0.99, \Lambda=0.01$ and $0$, respectively. The four curves represent four different values of $\gamma$. With $\gamma=-0.1$ and $\gamma=0$, these black holes have three real positive roots as shown in Fig.~\ref{fig:dsfig1}. The innermost zero is the black hole's inner horizon whereas the second and the third (outermost) zeroes are the black hole's event horizon and cosmological horizon respectively. For $\gamma=0.1$, there is only one horizon located at $r\approx37.5$. More interestingly with $\gamma=-0.2$, outside the horizon the metric function $f$ is always negative hence, the spacetime structure {\it outside} its horizon is similar to the {\it inside} spacetime structure of the standard Schwarzschild black hole. One observes that as $\gamma$ increases, the metric function develops its second and third node. Therefore, we expect that an extremal case $(f'(r_{h})=f(r_{h})=0)$ could exist at some point in the interval $0<\gamma<0.1$ as can be seen from Fig.~\ref{fig:dsfig1}. 

\subsection{Negative $\Lambda$}

\begin{figure}
        \centering
        \includegraphics[width=0.7\textwidth]{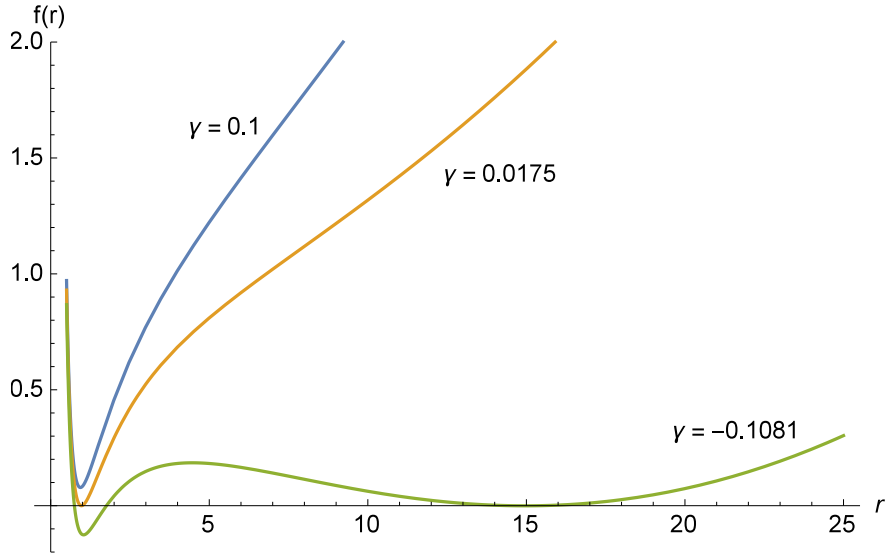}  
        \caption{The metric function $f(r)$ with differing values of $\gamma$.  For demonstration, we set $M=1, Q=0.99, \Lambda = -0.01, \epsilon = 0$.  For $\gamma > 0.0175$, the spacetime becomes regular with no horizon.  The spacetime contains extremal black hole when $\gamma = 0.0175, -0.1081$.  For $0.0175>\gamma>-0.1081$, we have non-extremal black hole spacetime.  When $\gamma < -0.1081$, the black hole becomes large with $r_{h}>\sqrt{3/|\Lambda|}$. } \label{fig1g}
\end{figure}

For negative cosmological constant $\Lambda$, the spacetime is asymptotically AdS.  To be specific, we set the mass, charge and cosmological constant term to be $M=1, Q=0.99, \Lambda=-0.01$ and consider the effect of $\gamma$ on the spacetime.  As shown in Fig.~\ref{fig1g}, changing $\gamma$ to large positive value could turn a black hole spacetime into a regular spacetime with no horizon but with naked singularity at $r=0$.  At approximately $\gamma=0.0175$, the black hole becomes extremal with inner regular spacetime behind the horizon due to the charge contribution.  For $0.0175>\gamma>-0.1081$, we have a small black hole~(with respect to $\sqrt{3/|\Lambda|}$).  At $\gamma=-0.1081$, the black hole becomes extremal again with regular inner region of spacetime behind the horizon.  In contrast to the extremal black hole in conventional gravity where charge contribution generates regular spacetime inside the horizon, this regular inner spacetime region originates from the massive-gravity {\it negative} $\gamma$ contribution.  For even more negative value of $\gamma<-0.1081$, the black hole becomes large.    
 
\section{Linear perturbations in electrovacuum}\label{sec:waveEq}

We shall now consider massive charged scalar field propagating in the background (\ref{BHmetric}). We assume there is no back-reaction of the scalar field onto the spacetime geometry. The evolution of charged scalar field can be described by the Klein-Gordon equation (\ref{scalar-eq}). Using the ansatz, $\Phi = e^{-i\omega t} \frac{\phi(r)}{r}Y(\theta,\varphi)$ with $Y(\theta,\varphi)$ the spherical harmonics and $A_\mu=\{A_0,0,0,0\}$, the scalar field equation becomes separable between the radial and angular part. The radial wave equation reads

\begin{align}
f \phi'' + f' \phi' + \left(\frac{1}{f}\left(\omega+qA_0\right)^2 - \frac{\ell(\ell+1)}{r^2} - \frac{f'}{r} - m_s^2 \right)\phi &= 0, \label{KG}
\end{align}

where $f'=df/dr$ and $-\ell(\ell+1)$ is the eigenvalue of the angular operator. This equation (\ref{KG}) can be recasted into the Schr\"{o}dinger-like form

\begin{align}
-\frac{d^2\phi}{dr_\ast^2} + \left[-\left(\omega+qA_0\right)^2 + f\left( m_s^2 + \frac{\ell(\ell+1)}{r^2} + \frac{f'}{r}     \right) \right]\phi &= 0, \label{KGtortoise}
\end{align}
where we have introduced the tortoise coordinate $r_\ast$
\begin{align}
\frac{dr_\ast}{dr} &= \frac{1}{f}.
\end{align}
If $\Lambda<0$, the tortoise coordinate is defined in the range $-\infty<r_\ast<\mathcal{C}$, where $r_\ast\rightarrow-\infty$ near the event horizon and at infinity $r_\ast\rightarrow \mathcal{C}$, where $\mathcal{C}$ is a positive constant. For $\Lambda>0$ case, $r_\ast\rightarrow -\infty,\infty$ as $r$ approaches outer event horizon $r_h$ and cosmological horizon $r_c$, respectively.

\section{QNMs of charged scalar in positive $\Lambda$ spacetime}\label{sec:dS}

%In this section, a massive charged scalar perturbation on a charged black hole in dRGT massive gravity with a positive cosmological constant $\Lambda>0$ will be studied. 

\subsection{Boundary condition}

In the vicinity of event horizon and cosmological horizon, general solution of (\ref{KGtortoise}) can be written down as

\begin{align}
\phi_{in} \sim \left\{ \begin{array}{lr}
  e^{-i\tilde{\omega} r_{\ast}},  \hspace{3.1cm} \mbox{ as $r\rightarrow r_h$} \\
  C_1 e^{-i\hat{\omega} r_{\ast}} + C_2 e^{i\hat{\omega} r_{\ast}}, \hspace{1.1cm} \mbox{ as $r\rightarrow r_c$}.
       \end{array} \right., \label{sol-in}
\end{align}

where $\tilde{\omega} \equiv \left(\omega+qA_h\right)$ and $\hat{\omega} \equiv \left(\omega+qA_c\right)$ for $A_h \equiv A_0(r_h)$ and $A_c \equiv A_0(r_c)$. Near the event horizon there is no outgoing wave whereas at the cosmic horizon there are both ingoing and outgoing modes. This is standard scattering problem in black hole physics.

For normal modes the effective potential in (\ref{KGtortoise}) is real, we can construct another linearly independent solution to (\ref{KGtortoise}) by taking complex conjugate of (\ref{sol-in}). We thus define $\phi_{out}=\phi^{\ast}_{in}$. Then we compute the Wronskian of these solutions by 
\begin{align}
W(\phi_{in},\phi_{out}) &= \phi_{in}\frac{d\phi_{out}}{dr_{\ast}} - \phi_{out}\frac{d\phi_{in}}{dr_{\ast}}.
\end{align}
Next, we obtain the following by evaluating the Wronskian at the event horizon and cosmological horizon,
\begin{align}
W&\bigg|_{r_{\ast}=-\infty} = 2i\tilde{\omega}, \\
W&\bigg|_{r_{\ast}=\infty} \hspace{2mm} = 2i\hat{\omega} \left( |C_1|^2 - |C_2|^2 \right).
\end{align}
Since Wronskian of linearly independent solutions must be a constant. We thus have 
\begin{align}
\frac{\tilde{\omega}}{\hat{\omega}} |T|^2 &= 1 - |R|^2,
\end{align}
where we have defined
\begin{equation}
C_1 = \frac{1}{T}, \qquad \frac{C_2}{C_1} = R.
\end{equation}
$|T|^2$ and $|R|^2$ are transmission and reflection coefficients respectively. One can see that if $|R|>1$, then we must have $\frac{\tilde{\omega}}{\hat{\omega}}<0$. This implies
\begin{align}
\frac{qQ}{r_c}<\omega<\frac{qQ}{r_h}, \label{SR-cond}
\end{align} 
where we choose $A_0=-Q/r$. If the frequency of the scalar field obeys this condition then its reflection amplitude will be greater than unity, we thus have superradiance effect. This result agrees with those found in \cite{Konoplya:2014lha} where superradiant of charged scalar field on RN-dS is investigated. Note that in the asymptotically flat limit, i.e., $r_c\rightarrow\infty$, this superradiant condition reduces to those in standard RN case \cite{Bekenstein:1973mi}. 

We shall now consider the quasinormal boundary condition. This is obtained by considering only outgoing mode at cosmological horizon. Thus we have 

\begin{align}
\phi_{in} \sim \left\{ \begin{array}{lr}
  e^{-i\tilde{\omega} r_{\ast}},  \hspace{2.8cm} \mbox{ as $r\rightarrow r_h$} \\
 e^{i\hat{\omega} r_{\ast}}, \hspace{3cm} \mbox{ as $r\rightarrow r_c$}.
       \end{array} \right., \label{QNMsBC}
\end{align}

The frequencies satisfying this boundary condition are called quasinormal frequency \cite{Zhidenko:2003wq,Berti:2009kk,Konoplya:2014lha}. This boundary condition implies that the frequencies $\omega$ are complex numbers. The scalar perturbation will be stable if Im$(\omega)<0$ (decaying). However if Im$(\omega)>0$ (growing), we have unstable modes. 

For the dS-type solution, we shall use the asymptotic iteration method (AIM) to compute the quasinormal modes. AIM has been firstly developed for obtaining solution of the second order ordinary differential equations \cite{AIM:2003}. Also, AIM is applied to compute the QNMs of Schwarzschild and Schwarzschild de-Sitter black hole \cite{Cho:2009cj}. Recently, the authors of \cite{Prasia:2016fcc} use AIM to study the QNMs of black holes in dRTG massive gravity.

%So far I have found two methods to derive an analytic formula. Both of them use the same idea of finding solution of Klein-Gordon equation in the near and the far region and then match those solutions in the intermediate region. I will follow both approaches and see which one gives us a better result. FYI, you can find more detail on these two methods in \cite{Crispino:2013pya} (section 3.c) \cite{Kanti:2014dxa} (section 3.1)

\subsection{Computation of QNMs using AIM}

To calculate the quasinormal frequencies using AIM, it is convenient to make a change of variable $r=1/x$. The radial part of the Klein-Gordon equation (\ref{KG}) becomes

\begin{align}
\phi'' + \frac{p'}{p}\phi' + \left[\frac{\left(\omega-q Q x\right)^2}{p^2} - \frac{1}{p}\left(\ell(\ell+1)+2Mx-2Q^2x^2+\frac{\gamma}{x}-\frac{2\Lambda}{3x^2} + \frac{m_s^2}{x^2}\right)\right]\phi &=0, \label{KG-inX}
\end{align}
where
\begin{align}
p &=  Q^2x^4 - 2Mx^3 + (1+\epsilon)x^2 + \gamma x - \frac{\Lambda}{3}.
\end{align}
In this section, $^{\prime}$ denotes a derivative with respect to $x$. It would be convenient to introduce \cite{Moss:2001ga,Cho:2009cj,Prasia:2016fcc}
\begin{align}
e^{i\omega r_*} &= (x-x_1)^{\frac{i\omega}{2\kappa_1}}(x-x_2)^{\frac{i\omega}{2\kappa_2}}(x-x_3)^{\frac{i\omega}{2\kappa_3}}(x-x_4)^{\frac{i\omega}{2\kappa_4}},
\end{align}
where $x_i=1/r_{i}$ for $i=1,2,3,4$ which represent each real roots of $f(r)$. The outer event horizon and cosmological horizon will be denoted by $x_1$ and $x_2$ whereas the inner event horizon and a negative real root are $x_3$ and $x_4$ respectively. We have also introduced the surface gravity which is defined as
\begin{align}
\kappa_i &= \left. \frac{1}{2}\frac{df}{dr}\right\rvert_{r\rightarrow r_i}.
\end{align}
For example, the surface gravity at the event horizon is denoted by $\kappa_1$. To scale out the divergent behaviour at the cosmic horizon, we define
\begin{align}
\phi(x) &= e^{i\omega r_*}u(x).
\end{align}
The wave equation (\ref{KG-inX}) therefore takes the following form 
\begin{align}
u'' + \frac{\left(p' - 2i\omega\right)}{p}u' - \frac{1}{p}\left[\ell(\ell+1)+2Mx-2Q^2x^2 +\frac{\gamma}{x}-\frac{2\Lambda}{3x^2}+\frac{m_s^2}{x^2} + \frac{qQx}{p}(2\omega-qQx) \right]u
\end{align}
In the absence of charge $q$ and mass $m_s$ of scalar field, this equation becomes similar to that of \cite{Prasia:2016fcc}. At the event horizon, the divergent behaviour is scaled out by taking
\begin{align}
u(x) &= (x-x_1)^{-\frac{i\omega}{\kappa_1}}\chi(x).
\end{align}
Finally, the radial equation becomes
\begin{align}
\chi''(x) &= \lambda_0(x) \chi'(x) + s_0(x) \chi(x),
\end{align}
with
\begin{align}
\lambda_0 &= -\frac{4 i \omega}{Q^2 (x-x_1) (x_1-x_2) (x_1-x_3) (x_1-x_4)}-\frac{p'-2 i \omega}{p},  \\
s_0 &= \frac{\ell(\ell+1)+2x(M-Q^2x)}{p} + \frac{2 m_s^2 + 3\gamma x + 2\Lambda}{3px^2} + \frac{qQx(2\omega-qQx)}{p^2}\nonumber \\ 
& -\frac{2 i \omega p'}{p Q^2 (x-x_1) (x_1-x_2) (x_1-x_3) (x_1-x_4)} 
  -\frac{4 \omega^2}{p Q^2 (x-x_1) (x_1-x_2) (x_1-x_3) (x_1-x_4)} \nonumber \\
&  +\frac{4 \omega^2}{Q^4 (x-x_1)^2 (x_1-x_2)^2 (x_1-x_3)^2 (x_1-x_4)^2}+\frac{2 i \omega}{Q^2 (x-x_1)^2 (x_1-x_2) (x_1-x_3) (x_1-x_4)}.
\end{align}

By differentiating the above equation with respect to $x$ for $n$ times, we obtain \cite{Prasia:2016fcc}
\begin{align}
\chi^{(n)} = \lambda_{n-2}\chi' + s_{n-2}\chi,
\end{align}
where the coefficients $\lambda_{n-2}$ and $s_{n-2}$ form a recurrent relation as
\begin{align}
\lambda_n &= \lambda'_{n-1} + \lambda_{n-1}\lambda_0 + s_{n-1}, \label{iter1} \\
s_n  &= s'_{n-1} + s_0\lambda_{n-1}. \label{iter2}
\end{align}
For sufficiently large $n$, the asymptotic behaviour implies
\begin{align}
\frac{s_n}{\lambda_{n}} = \frac{s_{n-1}}{\lambda_{n-1}}\equiv \beta,
\end{align}
where $\beta$ is a constant. The quasinormal frequencies $\omega$ can be found from the quantization condition \cite{Cho:2009cj}
\begin{align}
\lambda_n (x)s_{n-1}(x) &= \lambda_{n-1}(x)s_n(x). \label{quantcond}
\end{align}
To obtain the energy eigenvalues, each coefficients will be constructed in terms of their previous iteration via (\ref{iter1}) and (\ref{iter2}). This means each derivative of $\lambda$ and $s$ will also be determined. The quantization condition (\ref{quantcond}) will yield the expression for the energy eigenvalues. However, this becomes one of the main disadvantage of this method. Since at each step, one must calculate the derivative of $\lambda$ and $s$ of the previous iteration. This can be very time consuming and also affects the precision of the numerical calculation \cite{Cho:2009cj,Prasia:2016fcc}. To avoid this problem, an improved version of AIM has been proposed by the authors of \cite{Cho:2009cj}. The improved AIM overcomes the need to take the derivative at each iteration by expanding $\lambda_n$ and $s_n$ in a Taylor series around the point $\bar{x}$,
\begin{align}
\lambda_n(\bar{x}) &= \sum_{i=0}^{\infty} c^{i}_n (x-\bar{x})^i, \\
s_n(\bar{x}) &= \sum_{i=0}^{\infty} d^{i}_n (x-\bar{x})^i,
\end{align}
where $c^{i}_n$ and $d^{i}_n$ are the $i$-th Taylor coefficient's of $\lambda_n$ and $s_n$. Inserting these expressions into (\ref{iter1}) and (\ref{iter2}), we obtain
\begin{align}
c^{i}_n &= (i+1)c^{i+1}_{n-1} + d^i_{n-1} + \sum_{k=0}^{i}c^{k}_0 c^{i-k}_{n-1}, \label{iter1a} \\
d^{i}_n &= (i+1)d^{i+1}_{n-1} + \sum_{k=0}^{i}d^{k}_0 c^{i-k}_{n-1}. \label{iter2a}
\end{align}
The quantization condition (\ref{quantcond}) can be now expressed in terms of these new recursion relations
\begin{align}
d^{0}_{n}c^{0}_{n-1} - d^{0}_{n-1}c^{0}_{n} &= 0. \label{quantcond1}
\end{align}
Thus the improved AIM does not need the derivative operator. The quasinormal frequency can be obtained by solving a set of recursion relations above. The computational steps are as follows. First the coefficients $c^{i}_n$ and $d^{i}_n$ are computed via (\ref{iter1a}) and (\ref{iter2a}) starting from $n=0$ and iterating up to $n+1$ until it reaches the desired number of recursion. Then at each iteration $n$, the coefficients are determined with $i<N-n$, where $N$ is the maximum number of iterations, since the quantization condition (\ref{quantcond1}) contains only $i=0$. In this paper, for asymptotically de-Sitter solution, we will calculate the quasinormal mode of the dRGT charged black hole using the improved AIM.

\subsection{Results}

The QNMs of dRGT massive gravity de-Sitter black hole are calculated by Mathematica's notebook adopted from \cite{WadeNB}. In Table~\ref{tab:dsfig1}, we calculate the QNMs for massless charged scalar perturbation for various values of $\gamma$. The location of the black hole event horizon $x_1$ and the cosmic horizon $x_2$ change as $\gamma$ is varied. In this table, we show three sets of quasinormal frequencies $\omega$ distinguished by the expansion point $\bar{x}.$ We display three lowest modes of imaginary part for each fixed $\gamma$. In each cases, we find that the lowest mode becomes normal mode, i.e., zero imaginary part, while the real part is non-zero. Moreover, as the imaginary part increases (in magnitude), the real part of $\omega$ remains unchanged. The step of change in the imaginary part for the same real part is also found to be constant. These modes are the diffusive QNMs of the uncharged black hole being shifted by Coulomb energy when the scalar charge $q$ is turned on. We also observe that as $\gamma$ increases, the real part of $\omega$ decreases while the imaginary part of $\omega$ increases (in magnitude) when $\bar{x}=0.3x_1$. For the near-horizon solution ($\bar{x}=0.9x_1$), as $\gamma$ increases, both real part and imaginary part increase. For the expansion point $\bar{x}=0.3x_1$, these modes live outside the superradiant regime while all the results from another point ($\bar{x}=0.9x_1$) satisfy the superradiant condition (\ref{SR-cond}). Since we have found only QNMs with Im$(\omega)<0$, these modes are stable. 

However, unstable modes are found when the expansion point is $\bar{x}=\frac{x_1+x_2}{2}$. Note that the unstable QNMs obtained from the AIM converge relatively slowly when compare to the stable one. It can be seen from Table~\ref{tab:dsfig1} that for a given $\gamma$, only the most unstable modes live in the superradiant regime. Therefore not all the unstable modes discovered here are superradiant. For the QNMs of RN-dS \cite{Zhu:2014sya}, only $\ell=0$ modes are unstable and the nature of this instability is due to superradiance effect \cite{Konoplya:2014lha}. Moreover, even though all the unstable modes found in Refs \cite{Konoplya:2014lha} (for RN-dS) are superradiant but not all the superradiant modes are unstable.

A remarkable aspect of the results is the existence of three kinds of solutions categorized by $\bar{x}$, the near-$r_{h}$, the near-$r_{c}$, and the all-region solution.  The near-$r_{h}~(r_{c}$) uses $\bar{x}$ close to the event horizon~(cosmic horizon) at $x_{1}~(x_{2})$ respectively and the all-region solution uses $\bar{x}$ in the intermediate region between the two horizons.  In Table~\ref{tab:dsfig1}, the near-$r_{h}$~($r_{c}$) solution is given by AIM for $\bar{x}=0.9 x_{1}~(0.3 x_{1})$ respectively.  Each near-horizon solution are separated by the potential wall in the background and they have relatively smaller energies~(denoted by $\text{Re}(\omega)$) than the potential wall.  They are thus confined within the near-horizon regions with the wave function exponentially suppressed in the intermediate region where the potential wall dominates. These modes found in the near-horizon regions have identical real parts determined by quantity $qQ/2r_{h}~(qQ/2r_{c})$ as shown in Table~\ref{tab:dsfig1}. In fact, we observe that all the real parts for the near-horizon solutions in Table~\ref{tab:dsfig2}-\ref{tab:dsfig4} are equal to these factors. We can understand the shift in $\text{Re}(\omega)$~(i.e., energy) for the charged QNMs of these modes as the electric potential energy generated from the Coulomb interaction between the charged scalar and the charged black hole.  Table~\ref{tab:dsfig3} confirms this relationship, the value of $\text{Re}(\omega)$ is proportional to $q$ for a fixed $Q$ and it is equal to $qQ/2r_{h}~(qQ/2r_{c})$.  Therefore we can conclude that these near-horizon modes correspond to the diffusive QNMs of the uncharged black hole being shifted~(in the real parts) by the electric potential when the charge of the scalar field is turned on.  

On the other hand, the all-region solution has energy higher than the potential wall and thus their QNMs have much higher $\text{Re}(\omega)$ as shown in Table~\ref{tab:dsfig2}-\ref{tab:dsfig4} (exceptions are the unstable modes found in Table~\ref{tab:dsfig1} where the energies could become relatively small but still higher than the potential wall).  These QNMs in all-region solutions are those to be compared with values from the WKB method since WKB finds quantization condition from connecting solutions from the two regions around the maximum of the potential.

\setlength{\tabcolsep}{10pt}

\begin{table}[H]

\centering

\begin{tabular}{|c|c|c|c|c|c|}
\hline
\multicolumn{6}{|c|}{The QNMs calculated by improved AIM ($100$ iterations)} \\
\hline
$\gamma$ & $\omega(\bar{x}=0.3x_1)$ & $\omega(\bar{x}=0.9x_1)$ & $\omega(\bar{x}=\frac{x_1+x_2}{2})$ & $qQ/r_c$  & $qQ/r_h$  \\  \hline
\multirow{5}{*}{$-0.10$} & \multirow{3}{*}{0.045713 $+$ 5.52$\times10^{-14}$ $i$} & \multirow{3}{*}{0.091473 $-$ 7.06$\times10^{-15}$ $i$} & 0.027331 $+$ 0.009470$i$ & \multirow{5}{*}{0.091427} & \multirow{5}{*}{0.182945} \\
 & \multirow{3}{*}{0.045713 $-$ 0.035508$i$} & \multirow{3}{*}{0.091473 $-$ 0.064954$i$} & 0.067230 $+$ 0.013379$i$ &  &\\
 & \multirow{3}{*}{0.045713 $-$ 0.071017$i$} & \multirow{3}{*}{0.091473 $-$ 0.129908$i$} & 0.046779 $+$ 0.013992$i$ & & \\
 & & & 0.089034 $+$ 0.017426$i$ & & \\
 & & & 0.125727 $+$ 0.018862$i$ & & \\
 \hline
 
\multirow{3}{*}{$-0.05$} & 0.025576 $-$ 1.85$\times10^{-15} $ $i$ & 0.115275 $-$ 6.41$\times10^{-13}$ $i$ & \multirow{2}{*}{0.049368 $+$ 0.012092$i$} & \multirow{3}{*}{0.051152} & \multirow{3}{*}{0.230550} \\
 & 0.025576 $-$ 0.046854$i$ & 0.115275 $-$ 0.159514$i$ & \multirow{2}{*}{0.115124 $+$ 0.036522$i$}  & &\\
 & 0.025576 $-$ 0.093707$i$ & 0.115275 $-$ 0.319028$i$ & & &\\
 \hline
 
\multirow{3}{*}{$0.00$} & 0.015253 $+$ 4.96$\times10^{-15}$ $i$ & 0.130935 $-$ 1.07$\times10^{-12}$ $i$ & \multirow{2}{*}{0.021209 $+$ 0.010103$i$} & \multirow{3}{*}{0.030505} & \multirow{3}{*}{0.261869} \\
 & 0.015253 $-$ 0.050350$i$ & 0.130935 $-$ 0.236556$i$ & \multirow{2}{*}{0.106157 $+$ 0.042381$i$}  & &\\
 & 0.015253 $-$ 0.100700$i$ & 0.130935 $-$ 0.473111$i$ & & &\\
 \hline
 
\multirow{3}{*}{$0.05$} & 0.009612 $+$ 1.97$\times10^{-10}$ $i$ & 0.143604 $+$ 4.92$\times10^{-13}$ $i$ & \multirow{2}{*}{0.007692 $+$ 0.007593$i$} & \multirow{3}{*}{0.019223} & \multirow{3}{*}{0.287208} \\
 & 0.009612 $-$ 0.059340$i$ & 0.143604 $-$ 0.307075$i$ & \multirow{2}{*}{0.096506 $+$ 0.043602$i$} & &\\
 & 0.009612 $-$ 0.118680$i$ & 0.143604 $-$ 0.614150$i$ & & &\\
 \hline
 
\multirow{3}{*}{$0.10$} &  0.006619 $+$ 1.93$\times10^{-6}$ $i$  & 0.154574 $+$ 5.68$\times10^{-13}$ $i$ & \multirow{2}{*}{0.002261 $+$ 0.086304$i$} & \multirow{3}{*}{0.013178} & \multirow{3}{*}{0.309148} \\
& 0.006575 $-$ 0.074538$i$ & 0.154574 $-$ 0.373815$i$ & \multirow{2}{*}{0.086305 $+$ 0.042360$i$} &  &\\

& 0.006590 $-$ 0.149093$i$ & 0.154574 $-$ 0.747629$i$ & & &\\
 \hline

\end{tabular}
\caption{The QNMs for massless charged scalar perturbations of a charged dRGT black hole for $M=1$, $Q=0.5,\Lambda=0.01,q=0.99,\epsilon=0,\ell=0.$ Note that, number of iterations for the $\bar{x}=\frac{x_1+x_2}{2}$ case is $180.$}
\label{tab:dsfig1}
\end{table}

The effect of cosmological constant $\Lambda$ on the QNMs of charged black holes is shown in Table~\ref{tab:dsfig2}. For each fixed $\Lambda$, we show the three lowest modes of the quasinormal frequencies. With fixed $\Lambda$, we find a normal mode as the lowest possible mode. For the expansion point far from the black hole's horizon $\bar{x}=0.3x_1$, the real part and the imaginary part of the quasinormal frequencies increase as the cosmological constant increases. For the near-horizon point $\bar{x}=0.9x_1$, the real part and the imaginary part of $\omega$ decrease as $\Lambda$ increases. It is interesting that the imaginary part of QNMs with the same real part at each $\Lambda$ increases by a constant step for each evaluating point $\bar{x}=0.3 x_{1}, 0.9 x_{1}$. As discussed above, they are the diffusive modes of scalar field in uncharged black hole background being shifted by the Coulomb energy when the scalar charge $q$ is turned on. We also compute the QNMs by using the third-order WKB approximation (see Appendix \ref{app:A} for details). The results from WKB and AIM are compared where we have used another expansion point $\bar{x}=\frac{x_1+x_2}{2}$ \cite{2ndfootnote}. 
The results from two methods agree quite well with the difference only about $0.1\%$. Similar to the results displayed in Table~\ref{tab:dsfig1}, we find that some of theses frequencies satisfy superradiant condition (with $\bar{x}=0.9x_1$). In addition, we find no unstable mode since these modes exist with Im$(\omega)<0$.

In Table~\ref{tab:dsfig3}, the effect of scalar field charge $q$ on the QNMs of charged black holes is shown. In this case, the black hole event horizon locates at $x_1=0.652566$ and the cosmological horizon locates at $x_2=0.203419.$ For both set of quasinormal frequencies, the real part Re$(\omega)$ increases as the scalar charge $q$ increases. However, increasing $q$ does not affect the imaginary part of the quasinormal frequency. We notice that as $q$ increases, the real part of quasinormal frequencies is shifted up with the constant interval which is $0.009154$ for the first $\bar{x}$ and $0.029365$ for the second $\bar{x}$. The Coulomb shifts are simply $qQ/2r_{h}$ for $\bar{x}$ near the event horizon and $qQ/2r_{c}$ for $\bar{x}$ near the cosmic horizon. For the all-region solutions using $\bar{x}$ in the intermediate region, WKB and AIM both give the quasinormal frequencies which are in close agreement with each other. Some of these frequencies, i.e., mode with $\bar{x}=0.9x_1$, live in the superradiant regime where this can be seen by checking whether the real part of $\omega$ is satisfied by the condition (\ref{SR-cond}).

\begin{table}

\centering

\begin{tabular}{|c|c|c|c|c|}
\hline
\multicolumn{5}{|c|}{The QNMs calculated by improved AIM ($100$ iterations)} \\
\hline
$\Lambda$ & $\omega(\bar{x}=0.3x_1)$ & $\omega(\bar{x}=0.9x_1)$ & $\omega(\bar{x}=\frac{x_1+x_2}{2})$ & $3^{rd}$order WKB  $(n=0)$  \\  \hline
\multirow{3}{*}{$0.01$} & 0.002283 $+$ 2.27$\times10^{-9}i$ & 0.032466 $-$ 3.03$\times10^{-12}i$ &  &\\
 & 0.002283 $-$ 0.053227$i$ & 0.032466 $-$ 0.221712$i$ & 0.662813 $-$ 0.101124$i$ & 0.661603 $-$ 0.102322$i$ \\
 & 0.002289 $-$ 0.106455$i$& 0.032466 $-$ 0.443424$i$  & &\\
 \hline
 
\multirow{3}{*}{$0.05$} & 0.006169 $+$ 5.63$\times10^{-13}i$  & 0.031029 $+$ 2.94$\times10^{-12}i$  & &\\
 & 0.006169 $-$ 0.094880$i$ & 0.031029 $-$ 0.195734$i$ & 0.583884 $-$ 0.089959$i$ & 0.582789 $-$ 0.091289$i$ \\
 & 0.006169 $-$ 0.189760$i$ & 0.031029 $-$ 0.391468$i$  & &\\
 \hline
 
\multirow{3}{*}{$0.1$} & 0.009885 $-$ 6.11$\times10^{-15}i$  & 0.028898 $-$ 3.67$\times10^{-13}i$ & &\\
 &  0.009885 $-$ 0.102078$i$ & 0.028898 $-$ 0.155973$i$ & 0.469468 $-$ 0.072041$i$ & 0.468375 $-$ 0.073434$i$ \\
 & 0.009885 $-$ 0.204156$i$ & 0.028898 $-$ 0.311945$i$  & &\\
 \hline

\end{tabular}
\caption{The QNMs for charged scalar perturbations of a charged dRGT black hole for $M=1$, $Q=0.9,\gamma=0.02,q=0.1,m_s=0.2,\ell=2,\epsilon=0.$}
\label{tab:dsfig2}
\end{table}

%Note that, we cannot find a converged result in Table~\ref{tab:dsfig1} with this expansion point. Therefore it is not possible to compare the result from AIM with WKB for the case in Table~\ref{tab:dsfig1}.

%For fixed $\gamma$, as an angular index $\ell$ increases the real part of quasinormal frequencies also increases and the imaginary part becomes smaller (in absolute value). In addition, both real and imaginary part increase as $\gamma$ becomes more positive.

%In Table~\ref{tab:dsfig2} $M=1,Q=0.5,\gamma=-0.8,q=0,\ell=2$
%
%
%In Table~\ref{tab:dsfig3} $M=1,Q=0.5,\gamma=-0.8,m_s=0.1,\ell=2$
%
%In Table~\ref{tab:dsfig4} $M=1,Q=0.9,\gamma=0.02,\epsilon=0,m_s=0.2,\ell=2$
%
%In Table~\ref{tab:dsfig5} $M=1,\gamma=0.02,\epsilon=0,m_s=0.2,q=0.1,\ell=2$

\begin{table}

\centering

\begin{tabular}{|c|c|c|c|c|}
\hline
\multicolumn{5}{|c|}{The QNMs calculated by improved AIM ($100$ iterations)} \\
\hline
$q$ & $\omega(\bar{x}=0.3x_1)$ & $\omega(\bar{x}=0.9x_1)$  & $\omega(\bar{x}=\frac{x_1+x_2}{2})$ & $3^{rd}$order WKB  $(n=0)$ \\  \hline
\multirow{3}{*}{$0.1$} & 0.009154 $+$ 1.32$\times10^{-15}i$   & 0.029365 $-$ 3.56$\times10^{-12}i$   &  &\\
 & 0.009154 $-$ 0.102917$i$ & 0.029366 $-$ 0.164779$i$  & 0.494253 $-$ 0.076035$i$ & 0.493193 $-$ 0.077453$i$\\
 & 0.009153 $-$ 0.205834$i$ & 0.029366 $-$ 0.329558$i$  & & \\
 \hline
 
\multirow{3}{*}{$0.2$} & 0.018308 $+$ 1.77$\times10^{-15}i$    & 0.058731 $+$ 1.39$\times10^{-8}i$ & &  \\
 & 0.018308 $-$ 0.102917$i$ & 0.058734 $-$ 0.164776$i$ & 0.537927 $-$ 0.077051$i$ & 0.535105 $-$ 0.078125$i$ \\
 & 0.018308 $-$ 0.205834$i$& 0.058505 $-$ 0.329645$i$  & &   \\
 \hline
 
\multirow{3}{*}{$0.3$} & 0.027462 $+$ 8.78$\times10^{-15}i$    & 0.088096 $-$ 2.24$\times10^{-12}i$ & &   \\
 &  0.027462 $-$ 0.102917$i$ & 0.088096 $-$ 0.164779$i$ & 0.581910 $-$ 0.079698$i$ & 0.577703 $-$ 0.078755$i$ \\
 & 0.027462 $-$ 0.205834$i$ & 0.088097 $-$ 0.329558$i$  & & \\
 \hline

\multirow{3}{*}{$0.4$} & 0.036615 $-$ 3.00$\times10^{-15}i$     & 0.117462 $-$ 1.57$\times10^{-12}i$  & & \\
 &  0.036616 $-$ 0.102917$i$ & 0.117462 $-$ 0.164779$i$ & 0.625336 $-$ 0.083623$i$ & 0.620969 $-$ 0.079346$i$\\
 &  0.036616 $-$ 0.205834$i$ & 0.117462 $-$ 0.329558$i$  & & \\
 \hline

\end{tabular}
\caption{The QNMs for charged scalar perturbations of a charged dRGT black hole for $M=1$, $Q=0.9,\gamma=0.02,\Lambda=0.09,m_s=0.2,\ell=2,\epsilon=0.$ The black hole event horizon is at $x_1=0.652566$ and the cosmic horizon is at $x_2=0.203419.$}
\label{tab:dsfig3}
\end{table}

\setlength{\tabcolsep}{1.5pt}

\begin{table}

\centering

\begin{tabular}{|c|c|c|c|c|c|}
\hline
\multicolumn{6}{|c|}{The QNMs calculated by improved AIM ($100$ iterations)} \\
\hline
$m_s$ & $\omega(\bar{x}=0.3x_1)$ & $\omega(\bar{x}=0.9x_1)$   & $\omega_0(\bar{x}=0.6x_1)$ & $\omega_1(\bar{x}=0.6x_1)$ & $3^{rd}$order WKB  $(n=0)$   \\  \hline
\multirow{3}{*}{$0.00$}&$-$1.06$\times10^{-18}$ $-$ 0.341129$i$ & $-$3.08$\times10^{-6}$ $-$ 0.996799$i$ & 2.43$\times10^{-17}$ $-$ 0.725025$i$     &    &  \\
&6.78$\times10^{-19}$ $-$ 0.663874$i$ &5.99$\times10^{-6}$ $-$ 1.977313$i$ & $-$1.10$\times10^{-15}$ $-$ 1.248286$i$  & 1.597654 $-$ 0.429275$i$ & 1.594307 $-$ 0.429759$i$ \\
&$-$1.27$\times10^{-18}$ $-$ 1.035320$i$ &$-$0.000029 $-$ 2.967228$i$ & $-$1.94$\times10^{-14}$ $-$ 1.733901$i$ &  &\\
 \hline
 
\multirow{3}{*}{$0.25$} &$-$2.26$\times10^{-18}$ $-$ 0.346144$i$ &$-$2.93$\times10^{-6}$ $-$ 0.997271$i$ & 4.18$\times10^{-17}$ $-$ 0.778603$i$ &  &  \\
&$-$1.42$\times10^{-18}$ $-$ 0.677923$i$ &4.88$\times10^{-6}$ $-$ 1.976085$i$ & $-$1.62$\times10^{-15}$ $-$ 1.268964$i$  & 1.604779 $-$ 0.427158$i$ & 1.601012 $-$ 0.427723$i$  \\
&$-$1.25$\times10^{-19}$ $-$ 1.029516$i$ &$-$3.83$\times10^{-6}$ $-$ 2.984651$i$ & $-$2.99$\times10^{-14}$ $-$ 1.746501$i$  &   &\\
 \hline
 
\multirow{3}{*}{$0.50$} &1.72$\times10^{-18}$ $-$ 0.330507$i$ &$-$3.73$\times10^{-6}$ $-$ 0.996825$i$ & 4.13$\times10^{-16}$ $-$ 0.994794$i$&   &\\
&8.38$\times10^{-18}$ $-$ 0.622042$i$ &0.000015 $-$ 1.983883$i$ & $-$3.68$\times10^{-15}$ $-$ 1.277465$i$ & 1.625617 $-$ 0.421757$i$ & 1.621181 $-$ 0.421667$i$\\
&1.24$\times10^{-18}$ $-$ 1.001285$i$ &$-$6.81$\times10^{-6}$ $-$ 2.984021$i$ & $-$6.49$\times10^{-14}$ $-$ 1.781237$i$ &    &\\
 \hline
 
\end{tabular}
\caption{The QNMs for scalar perturbations of a charged dRGT black hole for $M=1$, $Q=0.5,\gamma=-0.8,\Lambda=0.08,q=0,\ell=2,\epsilon=1.984.$ The black hole event horizon is at $x_1=1.501496$ and the cosmic horizon is at $x_2=0.386506.$ Note that, the number of iterations for the $\bar{x}=0.9x_1$ case is $180$.}
\label{tab:dsfig4}
\end{table}

The QNMs for massive neutral scalar perturbations of a charged black holes in massive gravity background is displayed in Table~\ref{tab:dsfig4}. We show the results by varying the scalar field mass $m_s$ for $0,0.25$ and $0.50$. As in all previous tables, the QNMs are shown with three different evaluating points. Note that the all-region solution is computed using $\bar{x}=0.6x_1\approx\frac{x_1+x_2}{2}$. One distinguished feature of this table is, at $\bar{x}=0.6x_1$, we obtain two branches of quasinormal frequencies. In the first branch $\omega_0$, we show only three lowest modes of the quasinormal frequencies. In this branch, the real part of $\omega$ is zero which means that these modes are purely decayed or growing. Moreover increasing the scalar field mass $m_s$ slightly increases the imaginary part of $\omega$. For another branch where we dub as $\omega_1$, we find only one converged result for each fixed $m_{s}$. We see that $\omega_1$ is in the close agreement with the results obtained from WKB method. In this branch as $m_s$ increases, the real part of $\omega$ also increases monotonically and the imaginary part decreases. For the near-$r_h~(r_c)$ solution, the diffusive modes are also obtained. We display the lowest possible value of real part for the near-$r_{h}~(r_c)$ solution. Despite some of these real parts are not exactly zero but they are the actual diffusive modes. These non-zero real parts could be resolved by increasing the number of iteration. It should not be surprised that some of these modes acquire negative real part. This is because when $q=0$, the equation of motion (\ref{KG-inX}) has a symmetry under $\omega\to-\omega$. The effect of the scalar field mass on an imaginary part of $\omega$ is not straightforward for the near-$r_h~(r_c)$ solution. One final remark is, we find that these diffusive modes are shifted from the real axis when the scalar charge $q$ is switched on. The real parts will be shifted up with factor $qQ/2r_h(qQ/2r_c)$ as we increase $q$ in a similar way as found in Table~\ref{tab:dsfig3}.

\section{QNMs of charged scalar in negative $\Lambda$ spacetime}\label{sec:AdS}

In this section, we will consider the stability of black hole in the current massive gravity model when $\Lambda < 0$.  The boundary condition at the event horizon is the ingoing waves and at spatial infinity is zero.  In this case (\ref{KGtortoise}) reduces to
\beq
\frac{d^2\phi}{dr_\ast^2}= -\left(\omega+qA_0\right)^2 \phi, \label{adseq1}
\eeq
for $f(r)\approx 0$ near the horizon $r\simeq r_{h}$.  In this region, the scalar field takes the form
\beq
\phi(r) = A e^{-i(\omega +q A_{h})r_{*}}\equiv A e^{-i\tilde{\omega}r_{*}},
\eeq
where $A_{h}=k+V_{0}/r_{h}$.  Since
\beq
r_{*}=\int f^{-1}~dr \simeq \frac{1}{f'(r_{h})}\ln |r-r_{h}|,
\eeq
we can rewrite the field in the near-horizon region in the following form
\beq
\phi = f(r)^{-i\tilde{\omega}/4\pi T}\left( a_{0}+a_{1}(r-r_{h})+a_{2}(r-r_{h})^{2}+ ... \right),
\eeq
where $T$ is the Hawking temperature. 

As we approach $r\to \infty$, the equation of motion becomes
\beq
\frac{d^2\phi}{dr_\ast^2}= \left( m^{2}_s -\frac{2\Lambda}{3}\right) \frac{\Lambda r^{2}}{3} \phi. \label{adseq2}
\eeq
The solution of scalar field in the far away region is thus
\beq
\phi(r) = B r^{\alpha},
\eeq
where 
\beq
\alpha = -\frac{1}{2}\left( 1\mp \sqrt{9-\frac{12 m^{2}_s}{\Lambda}}\right). \label{alpha}
\eeq
We will choose only the plus sign since we need the field to vanish at infinity. On the other hand, there is a class of solution which also vanishes at infinity for the minus sign choice of (\ref{alpha}). It is simply required that $m^{2}_{s}/\Lambda > 2/3$.  Interestingly, for $\Lambda < 0,$ the possibility of negative mass square $m^{2}_{s}<0$ is also allowed as long as it satisfies the above requirement. 

In order to solve for the QNMs of the charged scalar in the massive gravity background, we rewrite the equation of motion (\ref{KGtortoise}) as the following
\beq
\frac{ \left[( w+\mu  (1-z))^2-f(z) \left(-z^{3}f'(z)+\ell(\ell+1) z^2+\tilde{m}^2\right)\right]\phi (z)}{f(z)}+z^2 \frac{\partial }{\partial z}\left(z^2 f(z) \frac{\partial \phi (z)}{\partial z}\right)=0,   \label{eomnum}
\eeq
where we define $z=r_{h}/r, w=\omega r_{h}, \tilde{m}=m_s r_{h}$.  With respect to the new coordinate, the physical region is $z \in [0,1]$, the infinity is at $z=0$ and the horizon is at $z=1$.  The electric potential is also expressed as 
\beq
qA_{0}=\mu\left( \frac{1}{r_{h}}-\frac{1}{r}\right),
\eeq
where $\mu = qQ$.  This choice of gauge picks horizon as the ground of the potential.  In order to calculate the QNM frequencies, we linearize the equation of motion with respect to $w$ by substitute into (\ref{eomnum})
\beq
\phi(z) =  e^{-i w r_{*}} S(z),
\eeq      
to obtain
\beqy
\frac{\left[ \left(2 \mu w (1-z)+(\mu  (1-z))^2\right)-f(z) \left(-z^3 f'(z)+\ell(\ell+1) z^2+\tilde{m}^2\right)\right]S(z)}{f(z)}&+&z^2 \frac{\partial }{\partial z}\left(z^2 f(z) \frac{\partial S(z)}{\partial z}\right)  \nonumber \\
&+&2 i w z^2 \frac{\partial S(z)}{\partial z}=0.   \label{leom1}
\eeqy

Alternatively, we can work in the Eddington-Finkelstein coordinates $v\equiv t + r_{*}$ and obtain the equation of motion for the scalar field,
\beq
-\Big(-z^{3}f'(z)+\ell(\ell+1) z^2+\tilde{m}^2+i \mu z^2\Big)\Psi (z)+z^2 \frac{\partial }{\partial z}\left(z^2 f(z) \frac{\partial \Psi (z)}{\partial z}\right)+2 i z^2 \Big(\mu (1-z)+w\Big)\frac{\partial \Psi (z)}{\partial z}=0,  \label{leom2}
\eeq
which is automatically linear in $w$.  Equations (\ref{leom1}) and (\ref{leom2}) are identical for $\mu = 0$.  Generically however even when $\mu$ is nonzero, we expect the quasinormal frequencies calculated from both equations to be the same.  We have numerically verified that the two equations of motion indeed give the same quasinormal frequencies.  

Expand for positive integer $N$ 
\beq
S(z) = \sum_{n=0}^{N}b_{n}T_{n}(2z-1),
\eeq
where $T_{n}$ is the Chebyshev polynomials of the first kind, we obtain the linear equation of coefficients $b_{n}$.  In the limit $N\to \infty$, the expansion will be exact due to the completeness of the orthonormal Chebyshev polynomials in domain $[-1,1]$.  To compute the quasinormal frequencies we adopt the spectral method by dividing the domain of interest $(2z-1) \in [-1,1]$ into a finite number of grid points and solve the system of linear equations of coefficients $b_{n}$.  The choice of grid points we adopt is the Gauss-Lobatto grid points 
\beq
z_{k}=\frac{1}{2}\left(1+\cos\left(\frac{k\pi}{N}\right)\right),
\eeq
where $k=0,1,..,N$.  The resulting system of linear equations is a generalized eigenvalue problem which can be solved to obtain the quasinormal frequencies $w$ for a given $N$.  The Mathematica code we used is adopted from Yaffe's method in Ref.~\cite{yaffe}    

\subsection{Small AdS black hole}

We will start with the set of physical parameters that gives small AdS black hole.  The parameter set we will use is a near-extremal black hole in conventional gravity, $M=1, Q=0.99, \Lambda =-0.01$.

\begin{figure}[h]
        \centering
        \includegraphics[width=0.8\textwidth]{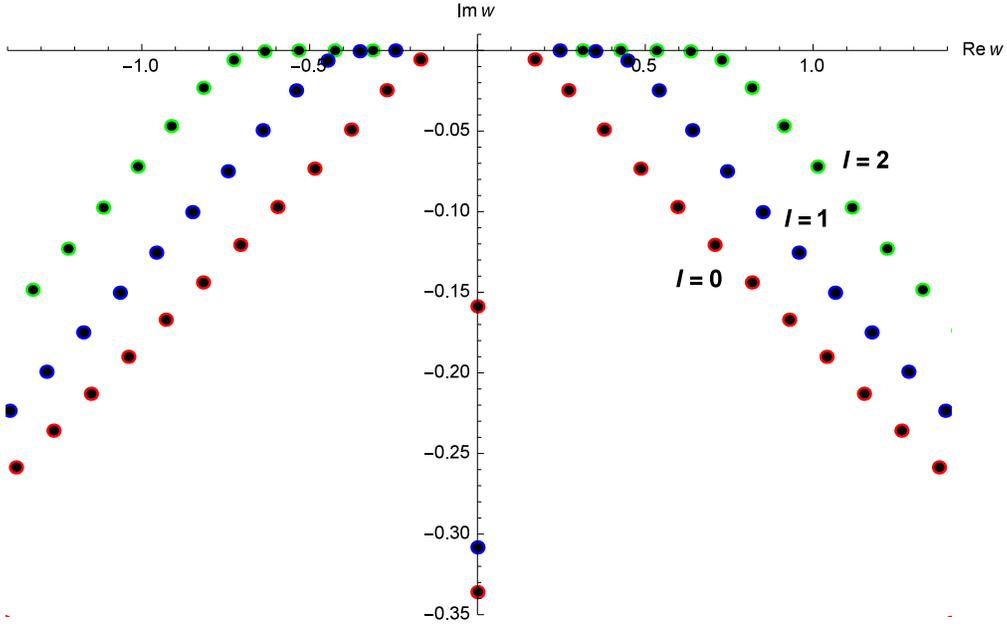}  
        \caption{The lowest QNMs $w=\omega r_{h}$ for $M=1, Q=0.99, \Lambda=-0.01, \gamma=0, \mu = 0, m_s=0, \ell=0,1,2$.  The red/black~(blue/black, green/black) dots are for $\ell=0~(1, 2)$ when $N=200/300$ respectively.  The convergence of the results are excellent as we can see no distinctive differences in the values of QNMs between $N=200$ and $N=300$. } \label{fig1}
\end{figure}

The QNMs of the massless neutral scalar are shown in Fig.~\ref{fig1} for the angular momentum states $\ell =0, 1, 2$.  Similar to the QNMs of fluctuations in the black brane geometry, the QNMs show approximate asymptotic linearity in both real and imagniary parts~(also previously shown in Ref.~\cite{Arav:2012ud}).  For $\ell=0, 1$, there are ``diffusive'' or ``hydrodynamic'' modes with zero real parts, another characteristic that is similar to the QNMs of the black brane spacetime.  

\begin{figure}[h]
        \centering
        \includegraphics[width=0.8\textwidth]{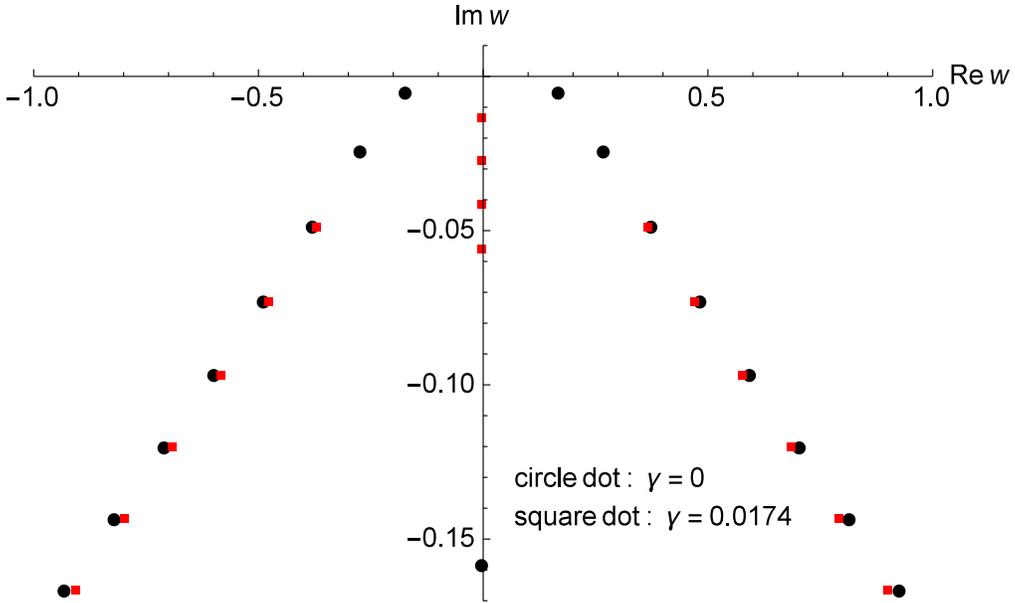}  
        \caption{The lowest QNMs $w=\omega r_{h}$ for $M=1, Q=0.99, \Lambda = -0.01, \mu = 0, \ell = 0, m_s =0$ for $\gamma = 0, 0.0174$.} \label{fig1a}
\end{figure}

\begin{figure}[h]
        \centering
        \includegraphics[width=0.8\textwidth]{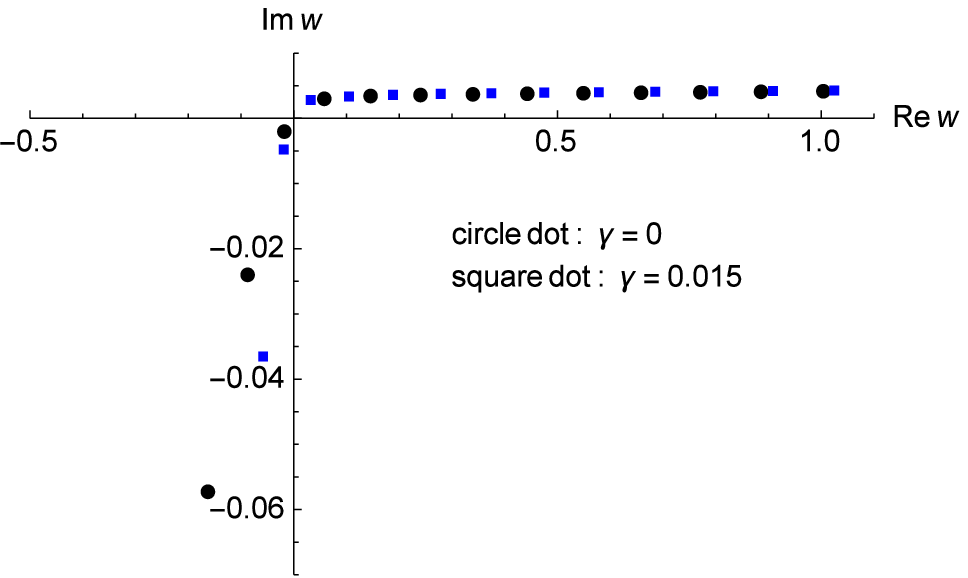}  
        \caption{The lowest QNMs $w=\omega r_{h}$ for $M=1, Q=0.99, \Lambda = -0.01, \mu = -2, \ell = 0, m_s =0$ for $\gamma = 0, 0.015$.} \label{fig2}
\end{figure}
The effect of massive gravity $\gamma$ parameter is shown in Fig.~\ref{fig1a}.  For diffusive modes with only imaginary parts, positive $\gamma$ generates more diffusive quasinormal frequencies with smaller values.  Analytic calculation of these modes is presented in section \ref{sec:Ana}.  Smaller diffusive QNMs imply longer relaxation time due to the massive gravity paramater $\gamma$.  For other QNMs with nonzero real parts, positive $\gamma$ slightly increases the slope of the asymptotic line, i.e., reducing the corresponding energy of each QNMs while keeping the imaginary part mostly unchanged.   

Figure \ref{fig2} shows unstable QNMs when the black hole has opposite charge to the scalar particle, i.e., $\mu=qQ < 0$ as we can see from the positive imaginary parts of $w$ for positive energy  Re$(w) > 0$.  The massive gravity parameter $\gamma$ does not affect the instability in a significant way as long as it does not change the background spacetime as mentioned above.  It only lowers the energy of the scalar in the unstable modes.  Using results of Ref.~\cite{Wang:2014eha}, after shifting electric potential at the horizon by $qQ/r_{h}$ to make ground voltage at the horizon, superradiant modes are the QNMs with Re$(\omega)<0$ for $\mu=qQ >0$.  This is equivalent to the QNMs with Re$(\omega)>0$ for $\mu <0$ since the equation of motion (\ref{KG}) depends on $(\omega+qA_{h})^{2}$ and has thus a symmetry under $\omega\leftrightarrow -\omega, q\leftrightarrow -q$.  Therefore, we can conclude that the unstable QNMs found are superradiant modes.

One possible interpretation of the unstable modes in the bulk is that the scalar will condensate around the black hole horizon resulting in superconducting layer outside the horizon~\cite{Gubser:2008px}.  Holographically, this would correspond to the superconducting phase of gauge theory on the AdS boundary.  From the viewpoint of holographic duality, the electric potential at the AdS boundary can be identified with chemical potential of the dual gauge matter living on the boundary.  We found that sufficiently large $|\mu|$ causes instability of the bulk scalar profile.  This could be interpreted as the instability of boundary dual matter phase to condensation.  It is possible that this is the condensation of scalar charged (quasi-)particles on the boundary rendering a superconducting phase~\cite{Gubser:2008px}.  Interestingly, the diffusive modes with zero real part disappear once the chemical potential $\mu$ is turned on.  

\subsection{Large AdS black hole}

For sufficiently large negative value of $\gamma$, e.g. $\gamma<-0.1081$ with $M=1,Q=0.99, \Lambda =-0.01$, the black hole turns into a {\it large} AdS black hole with $r_{h}\gtrsim R_{\rm AdS}=\sqrt{3/|\Lambda|}=17.32$.  Remarkably, the QNMs become almost purely diffusive for massless uncharged scalar~($m_s=0,\mu=0$) for $\ell =0,1,2$ as shown in Table~\ref{tab1}.  Other modes with nonzero real parts are non-converging at least up to $N=600$.  At $\gamma =-0.6$, the AdS black hole has a large size with $r_{h}/R_{\rm AdS}=10.3$.  The two lowest modes that can be obtained with reliable convengence for $N=600$ appear to be on the imaginary axis with very small real parts, the imaginary parts are almost identical between the states with $\ell=0,1,2$.  The oscillations are damped overcritically away with nearly zero frequencies.  Turning on the scalar charge, $\mu \neq 0$, does not change the results much, the first diffusive modes shift slowly to smaller imaginary values while the second diffusive modes disappear.  The convergence becomes very slow even at $N=600$.  For $\ell=0$ state, turning on $\mu$ shifts diffusive mode away from the imaginary axis.

\begin{table}[ht]
    
    \centering
    \begin{tabular}{|c|c|c|c|} 
    \hline
    $n$& $\ell=0$ & $\ell=1$ & $\ell =2$ \\ \hline
    1 & $2.28\times 10^{-8} - 78.571056 i$&$-3.51\times 10^{-8} -79.138264 i$  & $-2.42\times 10^{-8} -80.283881i$ \\ \hline
    2&$-0.0667-139.779252 i$&$0.039062 -140.051228 i$& $0.046945-140.4673596 i$\\
    \hline
    \end{tabular}
    \caption{The lowest QNMs $w=\omega r_{h}$ for $M=1, Q=0.99, \Lambda = -0.01, \mu = 0, m_s =0$ for $\gamma = -0.6$ and $\ell=0,1,2$.  The number of grid points is $N=600$.  The black hole is large with $r_{h}/R_{\rm AdS}=10.3$.}   
    \label{tab1}
\end{table}

It is curious that the QNMs of scalar perturbations in the large black hole in asymptotically AdS background in massive gravity model~(sufficiently large negative $\gamma$) become almost extinct with only few converging diffusive modes remain.  This phenomenon is purely massive gravity effect since for black hole with the same mass and charge~($M=1, Q=0.99$) in conventional gravity with $\gamma=0,\Lambda=-0.01$, the horizon radius $r_{h}=1.121\ll R_{\rm AdS}$, it is a small AdS black hole.  As shown in Fig.~\ref{fig1},\ref{fig1a} this small AdS black hole has series of QNMs of scalar perturbations.  Turning on massive gravity parameter $\gamma$ to a large negative value changes a small into a large AdS black hole.  As a result, most QNMs disappear with few diffusive modes survive.  

To complete the picture, we present QNMs of scalar field in the large AdS black hole spacetime in conventional gravity without massive graviton for $M=100, Q=9, \Lambda=-0.01$ in Table~\ref{tab2}~(we need to change the mass and charge since in Einstein gravity, the parameter set $M=1, Q=0.99$ will always give small AdS black hole or no black hole for $\Lambda<0$).  The convergence is very slow and only few reliably converging modes are found.  In contrast to large AdS black hole in massive gravity, the lowest QNMs have nonzero real parts and there are no diffusive modes found~(at least for this set of parameters).  QNMs of large AdS black hole in conventional gravity are  intrinsically different from the QNMs of large AdS black hole induced by pure massive gravity effect~(from otherwise small AdS black hole in Einstein gravity).      

It is possible that the very-slow convergence of QNMs is partially an artifact of the numerical method we are using in this article.  A different method, e.g. Frobenius mothod~\cite{Horowitz:1999jd} might reveal more converging QNMs for large AdS black hole even the one induced by massive gravity effect under consideration.  In order to check the converging aspect of our code with respect to the QNMs of large AdS black hole, we compare results with previous work~\cite{Horowitz:1999jd,Cardoso:2003cj}.  By numerically calculate the QNMs of scalar perturbation in large AdS black hole spacetime with $M=100, \Lambda=-0.01$, else $=0$~(this is large AdS black hole with $r_{h}/R_{\rm AdS}=2.11$), we found only the first converging modes~($n=1, w=\pm 9.333909-11.918457 i$) and the slowly converging second modes~($n=2, w\simeq \pm 15 -22 i$) even at $N=600$.  It demonstrates limitation on converging aspect of our code when applied to large AdS black hole with the full metric given by (\ref{metric}).      

\begin{table}[ht]
     
    \centering
    \begin{tabular}{|c|c|c|c|} 
    \hline
    $n$& $\ell=0$ & $\ell=1$ & $\ell =2$ \\ \hline
    1 & $\pm 9.208525 -11.853083 i$&$\pm 9.928185 -11.635843 i$  & $\pm 11.165467 -11.293674 i$ \\ \hline
    2&$-15.404565 -21.206122i$&$-16.178184 -21.188500 i$& $-16.933504 -20.997096 i$\\
     &$16.722404 -21.409218 i$ &$16.955786-21.151278 i$ &$16.369064 -21.117862 i$\\
    \hline
    \end{tabular}
    \caption{The lowest QNMs $w=\omega r_{h}$ for $M=100, Q=9, \Lambda = -0.01, \mu = 0, m_s =0$ for conventional gravity with $\gamma = 0$ and $\ell=0,1,2$.  The number of grid points is $N=600$.  The black hole is large with $r_{h}/R_{\rm AdS}=2.1$.}  
    \label{tab2}
\end{table}

\section{Analytic calculation of diffusive modes in massive gravity model}  \label{sec:Ana}

In this section, we will show that the QNMs of charged scalar field perturbation in the spacetime with nonzero $\Lambda$ and $\gamma$ are purely imaginary for $\ell = 0$.  This is the result of the boundary conditions on the scalar field at the far region $r\gg 1$.  \\

\underline{\bf AdS case}\\

First we will consider asymptotically AdS space with $\Lambda = -3/L^{2}$.  Start with the equation of motion (\ref{KG}) in the far region of the radial part $R(r)=\phi(r)/r$
\be
\left(\gamma r +\frac{r^{2}}{L^{2}}\right)R''(r)+\left( 3\gamma+\frac{4r}{L^{2}}\right)R'(r)+\left[ \frac{\omega^{2}}{\left( \gamma r +\frac{r^{2}}{L^{2}}\right)} - \left( \frac{\ell (\ell +1)}{r^{2}}+m_{s}^{2}\right)\right]R(r)=0,
\ee
where we have approximated $f(r)\simeq \gamma r + r^{2}/L^{2}$ in the far region.  In a new coordinate
\be
y\equiv 1+ \frac{r}{\gamma L^{2}},
\ee
the equation of motion in the far region can be rewritten as
\be
y(1-y)\frac{d^{2}R}{dy^{2}}+\left( 1-4y\right)\frac{dR}{dy}+\left[ \frac{\omega^{2}/\gamma^{2}}{y(1-y)}+\left( \frac{\ell (\ell+1)}{\gamma^{2}L^{2}(1-y)^{2}}+m_{s}^{2}L^{2}\right)\right].  \label{yeom}
\ee
To simplify the calculation we set $\ell =0$, the equation then has the following solutions
\be
R(y)=y^{-i\omegabar}(1-y)^{-1+\sqrt{1-\omegabar^{2}}}h(y),  \label{Rsol0}
\ee
where
\be
h(y)=A ~{}_{2}F_{1}(a,b,c;y)+B(-y)^{2i\omegabar} ~{}_{2}F_{1}(\omegabar \to -\omegabar),  \label{hsol}
\ee
when $_2F_1(a,b,c;y)$ is the hypergeometric function with 
\bea
a&&=-i\omegabar+\frac{1}{2}+\sqrt{1-\omegabar^{2}}-\frac{1}{2}\sqrt{9+4m^{2}},  \\
b&&=-i\omegabar+\frac{1}{2}+\sqrt{1-\omegabar^{2}}+\frac{1}{2}\sqrt{9+4m^{2}},
\eea
and $c=1-2i\omegabar$.
We have used dimensionless parameters $\omegabar\equiv \omega/\gamma, m\equiv m_{s}L$.  $A,B$ are constants to be determined by the boundary conditions.  Note that the second term on the RHS of (\ref{hsol}) is exactly the symmetric $\omegabar \to -\omegabar$ of the first term.  For consideration of the QNMs, it suffices to consider only the first term of the solution.  

In the far region $r\gg 1$, we can use the Euler's transformation on the hypergeometric function to obtain
\be
R(y)\sim \left( \frac{r}{\gamma L^{2}}\right)^{-\frac{3}{2}+\frac{3}{2}\sqrt{1+4m^{2}/9}}\frac{\Gamma(c)\Gamma(b-a)}{\Gamma(b)\Gamma(c-a)}, \label{rsol}
\ee
where we have used the identity
\be
{}_{2}F_{1}(a,c-b,c;1)=\frac{\Gamma(c)\Gamma(b-a)}{\Gamma(c-a)\Gamma(b)}. \notag
\ee
For $m^{2}<0,$ the exponent of $r/\gamma L^{2}$ in (\ref{rsol}) is negative, the solution is vanishing at infinity and there is no requirement on the $\omegabar$.  For $m^{2}\geq 0$ on the other hand, the exponent is positive and we need the poles of Gamma function to suppress the solution at infinity, i.e.,
\be
-i\omegabar+\frac{1}{2}\pm \sqrt{1-\omegabar^{2}}+\sqrt{9+4m^{2}}=-N,  
\ee
where $N$ is non-negative integer.  This leads to the QNMs
\be
\omega = -i \gamma \frac{\left(N+\frac{1}{2}(1+\sqrt{9+4m^{2}})\right)^{2}-1}{2\left(N+\frac{1}{2}(1+\sqrt{9+4m^{2}})\right)}, \label{diffqnm}
\ee
for $N=0,1,2,...$.  
However, the QNMs also need to satisfy the boundary condition at small $r$.  By using Pfaff's transformation
\be
{}_{2}F_{1}(a,b,c;z)=(1-z)^{c-a-b}{}_{2}F_{1}(c-a,c-b,c;z),
\ee
the far-region solution for small $r$ can be expressed as
\bea
R(y)&&\sim \left( \frac{r}{\gamma L^{2}}\right)^{-1+\sqrt{1-\omegabar^{2}}}\left( 1+\frac{r}{\gamma L^{2}}\right)^{-i\omegabar}\left( \frac{-r}{\gamma L^{2}}\right)^{c-b-a}{}_{2}F_{1}(c-a,c-b,c;y), \notag \\
&&\sim \left( \frac{r}{\gamma L^{2}}\right)^{-1-\sqrt{1-\omegabar^{2}}}\frac{\Gamma(c)\Gamma(2\sqrt{1-\omegabar^{2}})}{\Gamma(a)\Gamma(b)}.
\eea
For the solution to be vanishing, it is required that $b=-N$~($a=-N$ is not consistent with (\ref{diffqnm})), giving again the condition (\ref{diffqnm}).  There is a distinct difference between an asymptotically AdS space with and without the massive gravity effect $\gamma$.  The QNMs we found here are diffusive in nature with pure imaginary values which exist only when $\gamma$ is nonzero.  Numerical analysis confirms these diffusive modes as shown in Fig.~\ref{fig1a}.  Existence of small black hole in the small $r$ region changes the value of these diffusive QNMs by a small quantity, however, in addition to generating other possible vibrating modes of the QNMs.  These other possible modes already exist in a small AdS black hole in conventional Einstein gravity.
\\

\underline{\bf dS case}\\

A similar calculation can be performed in the asymptotically dS case with $\Lambda > 0$.  The equation of motion in the new coordinate $y\equiv 1-r/\gamma L^{2}$ can be written as
\be
y(1-y)\frac{d^{2}R}{dy^{2}}+\left( 1-4y\right)\frac{dR}{dy}+\left[ \frac{\omega^{2}/\gamma^{2}}{y(1-y)}+\left( \frac{\ell (\ell+1)}{\gamma^{2}L^{2}(1-y)^{2}}-m_{s}^{2}L^{2}\right)\right]. 
\ee
This is $L^{2}\to -L^{2}$ of (\ref{yeom}), so it has exactly the same solution  as (\ref{Rsol0}),(\ref{hsol}) with $m^{2}\to -m^{2}$.  Certainly, the far region is different since the spacetime boundary now becomes the cosmic horizon at $r_{c}=\gamma L^{2}$.  Similar to the AdS case, the far-region solution for $\ell = 0$ in the small $r$ limit~($y\to 1$) takes the following form
\bea
R(y)&&\sim \left( \frac{r}{\gamma L^{2}}\right)^{-1-\sqrt{1-\omegabar^{2}}}{}_{2}F_{1}(c-a',c-b',c;1), \notag \\
&&\sim \left( \frac{r}{\gamma L^{2}}\right)^{-1-\sqrt{1-\omegabar^{2}}}\frac{\Gamma(c)\Gamma(2\sqrt{1-\omegabar^{2}})}{\Gamma(a')\Gamma(b')},  \label{srdSsol}
\eea
where $a'=a(m^{2} \to -m^{2}), b'=b(m^{2} \to -m^{2})$ respectively.  Therefore, we have the QNMs given by
\be
\omega = -i \gamma \frac{\left(N+\frac{1}{2}(1+\sqrt{9-4m^{2}})\right)^{2}-1}{2\left(N+\frac{1}{2}(1+\sqrt{9-4m^{2}})\right)}, \label{diffqnm1}
\ee
for $N=0,1,2,...$ in order to make the far-region solution vanishes at small $r$.

The QNMs considered will certainly be modified by the presence of black hole, either by developing real parts and new QNMs as well as shifting the imaginary values of the original diffusive modes unique to the massive gravity model.  As is argued in Ref.~\cite{Uchikata:2011zz}, turning on the electric potential at the horizon should simply shift the real parts of QNMs by $qA_{h}=qQ/r_{h}$ for the asymptotically AdS case.  In the dS case, the results in Table~\ref{tab:dsfig1}-\ref{tab:dsfig3} show that shifts of the real parts of QNMs in the near-horizon regions are actually $qQ/2r_{h}~(qQ/2r_{c})$ for the near-$r_{h}~(r_{c})$ solutions respectively.  In the presence of massive gravity parameter $\gamma$, shifts in the real parts also depend on $\gamma$ as shown in Table~\ref{tab:dsfig1}.

\section{Conclusions}\label{sec:conclude}

In this paper, we have studied the effect of massive charged scalar perturbations on charged black hole spacetime in dRGT massive gravity. Notable effects of massive gravity are generation of cosmological constant term and the linear term in the metric (\ref{metric}) from combination of massive graviton mass, cubic and quartic graviton interactions and the fiducial metric.  Physically, only the fiducial metric determines the linear $\gamma r$ term in the sense that it will be zero if $c$ is vanishing.  A physical interpretation of the fiducial metric is an extra dimensional pullback from the bulk metric~\cite{Gabadadze:2015goa} or from the second site of the two-site theory~\cite{ArkaniHamed:2002sp}. We have explored the effect of the $\gamma$ term on the spacetime structure. It is found that the spacetime structure is very sensitive to the value of $\gamma$. For the de Sitter~(dS) case, the metric (\ref{metric}) have three positive real roots associated with the Cauchy horizon, the event horizon and the cosmological horizon respectively.  At some fixed value of $\gamma$, we obtain an extremal charged black hole where the Cauchy and event horizon of black hole coincide. For the AdS case, one can have standard charged AdS black hole, extremal charged AdS black hole or even regular spacetime with no horizon depending on the value of $\gamma$.

In section \ref{sec:dS}, the QNMs of black hole in asymptotically dS space in dRGT model are computed. The numerical scheme called asymptotic iteration method (AIM) \cite{AIM:2003} is applied for calculating the quasinormal frequencies. We find unstable modes for the lowest angular harmonic index $\ell=0$. Some of these unstable modes live in the superradiant regime. However not all the superradiant modes are unstable. On the other hand, we find no evidence of any instabilities for $\ell=2$. All the dS black holes we investigate in this case appear to be stable under small perturbation. Since the perturbation decays with time, therefore these scalar modes do not suffer from the superradiant instability. The QNMs for the all-region solutions are also computed via the third-order WKB approximation and are in good agreement with the results obtained from AIM. With $q=0$, the quasinormal frequencies with vanishing real part (the diffusive mode) are discovered. We find that as $q$ increases, the real part of $\omega$ is shifted up by $qQ/2r_{h}~(qQ/2r_{c})$ for these modes.  They correspond to the near-event~(cosmic)-horizon solutions respectively.  In addition, when $q$ is non-zero, the normal mode with zero imaginary part are found as the lowest possible mode. Finally, it is found that the black holes become more stable as $\gamma$ gets larger.

In section \ref{sec:AdS}, the QNMs of black hole in asymptotically AdS space in dRGT massive gravity are explored.  For a small black hole in negative cosmological constant spacetime, the QNMs have asymptotic linear dependence on the mode number $n$.  This can be shown using monodromy method and other approximation schemes~(see e.g. \cite{Berti:2009kk} and references therein) for asymptotically AdS space in conventional Einstein gravity.  In massive gravity model considered here, the linearity persists as long as the massive gravity effect does not regulate or alternate the black hole spacetime.  Charged perturbation of scalar field in the charged AdS black hole background could become unstable with positive imaginary parts of QNMs~(Fig.~\ref{fig2}). Massive gravity effects reduce the energy of these unstable modes but leave the characteristic time almost unchanged.  In holographic viewpoint, the gauge theory dual of these situations is possibly the condensation of charged scalar (quasi-)particles resulting in superconducting phase. 

Analytic calculation of the QNMs in the spacetime with massive gravity parameter $\gamma$ in Section~\ref{sec:Ana} shows that the QNMs form diffusive tower with the size proportional to $\gamma$ for $\ell =0$.  In contrast to empty AdS where all the QNMs are normal modes, the linear term $\gamma r$ in the metric induces a pseudo-horizon at $r=0$ in the asymptotically far region rendering all QNMs imaginary.  Numerical results in both dS and AdS cases confirm existence of these modes albeit deformed by presence of a black hole in the background.  
 
As a possible extension of this work, a time-domain analysis of linear charged scalar perturbation on charged dRGT black hole background is needed to confirm the frequency-domain stability presented in this paper. On the other hand, since we have shown that the AdS black holes are superradiantly unstable. Therefore it would be an interesting task to investigate the end-point of this instability. Many works have suggested that a hairy black hole could be the end point of superradiant instability \cite{Dolan:2015dha,Sanchis-Gual:2016tcm,Bosch:2016vcp}. This would require the analysis of a fully coupled system of dRGT massive gravity couples with a bosonic field. 

\appendix

\section{WKB APPROXIMATION}
\label{app:A}

In order to study QNMs of black holes, since the equation of linear perturbation of a black hole can be recasted into the Schr\"odinger-like form, one may apply WKB approximation to estimate possible modes of a black hole. In conventional quantum mechanics, the essence of the WKB method involves considering approximate solutions to each asymptotic region and matching them together to obtain an approximated QNMs of the black hole. The WKB approach makes use of a Schr\"odinger-like differential equation of the
following form \cite{Schutz:1985zz,Iyer:1986np,Konoplya:2003ii},
\begin{align}
\frac{d^2\phi}{dr_*^2}+\mathcal{Q}(r_*)\phi=0. \label{WKBmastereq}
\end{align}
Comparing this with (\ref{KGtortoise}) we can find the corresponding $\mathcal{Q}$ as
\begin{align}
\mathcal{Q} = \left(\omega+q A_0\right)^2-f\left(m^2_s+\frac{\ell\left(\ell+1\right)}{r^2}+\frac{f'}{r}\right). \label{Qform}
\end{align}
In the Schr\"{o}dinger's equation language, $\mathcal{Q}$ is equivalent to $2m\left(E-U\right)/\hbar^2$ where $m$ is mass, $E$ is energy, and $U$ is potential of a one-dimensional system. Despite the probably complex form of $\mathcal{Q}$, the property of $\mathcal{Q}$ around its extremum $r_{*0}$ can be approximated to be that of a parabola. To this end, we can perform a Taylor expansion of $\mathcal{Q}$ around its extremum point as follows,
\begin{align}
\mathcal{Q}(r_*)=\mathcal{Q}(r_{*0})+\frac{1}{2}\mathcal{Q}''(r_{*0})\left(r_*-r_{*0}\right)^2+ \ldots, \qquad \mathcal{Q}'(r_{*0})=0,
\end{align}
where a prime denotes a derivative with respect to $r_*$. From now on, we will use the following short-hands notation,
\begin{align}
\mathcal{Q}_0 \equiv \mathcal{Q}(r_{*0}),\qquad \mathcal{Q}''_0 \equiv \mathcal{Q}''(r_{*0}), \qquad \text{and so on.}
\end{align}
Through this expansion, it is clear that we can estimate a solution around $r_{*0}$ to be that of a system whose $\mathcal{Q}$ (or the corresponding potential $U$) is parabola. By requiring the boundary condition for QNMs (near-horizon field goes into the black hole and the field at $r\to\infty$ goes outwards to $\infty$), it is possible to obtain the following matching condition from the approximate solution,
\begin{align}
\frac{i\mathcal{Q}_0}{\sqrt{2\mathcal{Q}''_0}}-i\bar{\Lambda}-\Omega=\left(n+\frac{1}{2}\right), \label{WKBmatcheq}
\end{align}
where 
\begin{align}
\bar{\Lambda}&=\frac{1}{\left(2\mathcal{Q}''_0\right)^{1/2}} \left[ \frac{1}{8}\left(\frac{\mathcal{Q}^{(4)}_0}{\mathcal{Q}''_0}\right)\left(\frac{1}{4}+\alpha^2\right)-\frac{1}{288}\left(\frac{\mathcal{Q}'''_0}{\mathcal{Q}''_0}\right)^2\left(7+60\alpha^2\right) \right],
\\
\Omega&=\frac{\alpha}{2\mathcal{Q}''_0}\left[\frac{5}{6912}\left(\frac{\mathcal{Q}'''_0}{\mathcal{Q}''_0}\right)^4\left(77+188\alpha^2\right)-\frac{1}{384}\left(\frac{\mathcal{Q}'''^2_0 \mathcal{Q}^{(4)}_0}{\mathcal{Q}''^3_0}\right)\left(51+100\alpha^2\right)+\frac{1}{2304}\left(\frac{\mathcal{Q}^{(4)}_0}{\mathcal{Q}''_0}\right)^2\left(67+68\alpha^2\right)\right.
\\
&\qquad \left. \frac{1}{288}\left(\frac{\mathcal{Q}'''_0 \mathcal{Q}^{(5)}_0}{{\mathcal{Q}''_0}^2}\right)\left(19+28\alpha^2\right)-\frac{1}{288}\left(\frac{\mathcal{Q}^{(6)}_0}{\mathcal{Q}''_0}\right)\left(5+4\alpha^2\right)\right],
\\
\alpha&\equiv n+\frac{1}{2}, \quad n\in \left\{0,1,2,...\right\} \text{ for } \text{Re}(\omega)>0. \label{WKBdefinition}
\end{align}
The fundamental mode is represented by $n=0$. Particularly, the first term on the left-hand side of (\ref{WKBmatcheq}) corresponds to the first-order WKB approximation \cite{Schutz:1985zz}, the second term and the third term correspond to the second-order and third-order WKB approximations respectively \cite{Iyer:1986np,Konoplya:2003ii}.

In the Schwarzschild case, $\mathcal{Q}$ can be expressed as $\mathcal{Q}=\omega^2-V(r_*)$, $V$ is an $\omega$-independent function, which simply makes the evaluation of the second derivative of $\mathcal{Q}$ to be relatively easy. 
However, in general, each order of derivative of $\mathcal{Q}$ depends on $\omega$ which makes the matching condition cumbersome to deal with. In our study, We use the following techniques. First, we rewrite (\ref{WKBmatcheq}) as follows,
\begin{align}
i \mathcal{Q}_0 = \sqrt{2\mathcal{Q}''_0}\left(i\bar{\Lambda}+\Omega+\left(n+\frac{1}{2}\right)\right).
\end{align}
We then substitute $Q_0$ by using (\ref{Qform}) as follows,
\begin{align}
\left.\left[\left(\omega+q A_0\right)^2-f\left(m^2_s+\frac{\ell\left(\ell+1\right)}{r^2}+\frac{f'}{r}\right)\right]\right|_{r_0} =\frac{\sqrt{2\mathcal{Q}''_0}}{i}\left(i\bar{\Lambda}+\Omega+\left(n+\frac{1}{2}\right)\right). \label{WKBmatcheq2}
\end{align}
In order to approximate the quasinormal frequencies, we will perform an iteration using (\ref{WKBmatcheq2}). Given a random value of $\omega_0$, we compute for $r_0(\omega_0)$ which minimizes $\mathcal{Q}(\omega_0)$ then we find $\omega_1$ from the followings,
\begin{align}
\left(\omega_1+q A_0(r_0)\right)^2-\left.\left[f\left(m^2_s+\frac{\ell\left(\ell+1\right)}{r^2}+\frac{f'}{r}\right)\right]\right|_{r_0} =\frac{\sqrt{2\mathcal{Q}''_0(r_0,\omega_0)}}{i}\left(i\bar{\Lambda}(r_0,\omega_0)+\Omega(r_0,\omega_0)+\left(n+\frac{1}{2}\right)\right).
\end{align}
Furthermore, we find successive $\omega_i$ iteratively via the similar equation,
\begin{align}
\left(\omega_{i+1}+q A_0(r_0)\right)^2-\left.\left[f\left(m^2_s+\frac{\ell\left(\ell+1\right)}{r^2}+\frac{f'}{r}\right)\right]\right|_{r_0} =\frac{\sqrt{2\mathcal{Q}''_0(r_0,\omega_i)}}{i}\left(i\bar{\Lambda}(r_0,\omega_i)+\Omega(r_0,\omega_i)+\left(n+\frac{1}{2}\right)\right),
\end{align}
where $r_0=r_0(\omega_i)$ in this case. 
The iteration is performed until the difference between successive frequencies is less than $1\%$ and we take the frequency from the last iteration to be our approximate quasinormal frequency. 

In spite of the simple procedures, the WKB approximation is claimed to yield satisfactory results when the azimuthal number $\ell$ is greater than $n$. Moreover, since we consider two kinds of black hole solutions; dS and AdS black holes, the matching condition used here only corresponds to the dS case. In the dS case, the field at $r\to\infty$ tends to propagate outwards as a plane wave while this is not the case in the AdS case. The matching techniques in Ref. \cite{Schutz:1985zz,Iyer:1986np,Konoplya:2003ii} only utilize the boundary condition of a field propagating outwards, thus corresponds to the property of far-field limit in dS geometry.

\acknowledgments

We would like to thank Napat Poovuttikul for useful discussions.  L.T. is supported by the National Research Foundation of Korea (NRF) grant funded by the Korea government (MSIP) (No.2016R1C1B1010107). S.P. is supported by the Theoretical and Computational Science (TaCS) Center under ComputationaL and Applied Science for Smart Innovation Cluster (CLASSIC), Faculty of Science, KMUTT.

\bibliographystyle{apsrev}

\begin{thebibliography}{20}
\bibitem{1stfootnote}
There is a possibility that $\epsilon$ is extremely small but not zero, its size is constrained by the flatness of empty space in the spacetime, for simplicity we will simply assume $\epsilon=0$ here.
\bibitem{2ndfootnote} 
We will not compare the results from AIM and WKB for the data in Table~\ref{tab:dsfig1}. This is because the accuracy of the third-order WKB approximation is good for $\ell>n$, where $n$ is a fundamental mode. In addition, for a given $n$, the accuracy of the WKB method increases with $\ell$.
\end{thebibliography}

\end{document}